\documentclass[aps,preprint,nofootinbib,preprintnumbers]{revtex4}
\usepackage{amsmath,amstext,amssymb}
\usepackage{comment,url}
\usepackage{rotating}
\usepackage{graphicx,subfigure}
\newcommand{\beq}{\begin{equation}}
\newcommand{\eeq}{\end{equation}}

\begin{document}

% title page
%\preprint{\today}

\author{Stefan Janiszewski}
\email{stefanjj@uvic.ca}
\affiliation{Department of Physics and Astronomy, University of Victoria, Victoria, BC, V8P 5C2, Canada}
\author{Matthias Kaminski}
\email{mski@ua.edu}
\affiliation{Department of Physics and Astronomy, University of Alabama, Tuscaloosa, AL 35487, USA}

\title{
Quasinormal modes of magnetic and electric black branes versus far from equilibrium anisotropic fluids}

\begin{abstract}
Detailed knowledge about equilibration processes is of interest for various fields of physics, including heavy ion collision experiments and quantum quenched condensed matter systems. 
We study the approach to equilibrium at late times within two types of strongly coupled thermal systems in $3+1$ dimensions: systems in the presence of (i) a non-zero charge density, or (ii) a magnetic field at vanishing charge density. 
Utilizing the gauge/gravity correspondence, we map the aforementioned problem to the computation of quasinormal frequencies around two particular classes of black branes within the Einstein-Maxwell theory.
We compute (i) the tensor and vector quasinormal modes of Reissner-Nordstr\"om black branes and (ii) the scalar, as well as tensor quasinormal modes of magnetic black branes. 
Some of these quasinormal modes correspond to the late-time relaxation of the above systems after starting with initial pressure anisotropy. We provide benchmarks which need to be matched at late-times by all holographic thermalization codes with the appropriate symmetries.
\end{abstract}

\date{\today}

\maketitle

\tableofcontents

%%%%%%%%%%%%%%%%%%%%%%%%%%%%%%%%%%%%%%%
\section{Introduction} \label{sec:introduction}

\begin{figure}[h]
\includegraphics[height=7cm]{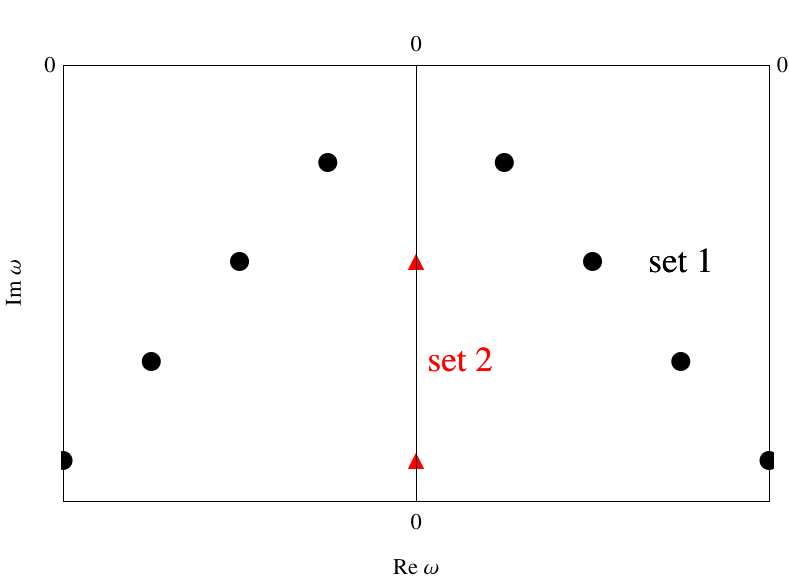}
\caption{\label{fig:QNMSets}
{\it Schematic plot of QNM sets.} Two distinct sets of QNMs of $h_{xy}$ fluctuations around Reissner-Nordstr{\"o}m black branes are shown. Set 1 (black dots) is the set which was found already in the case of a black brane with vanishing charge~\cite{Starinets:2002br}. Set 2 (red triangles) is a set of QNMs wandering up the imaginary frequency axis as the black brane charge is pushed towards its extremal value. Thus, at large charge the purely imaginary QNMs of set 2 will dominate the late-time behavior of the dual field theory plasma.
}
\end{figure}

Systems far from equilibrium have been largely inaccessible for a long time despite being abundant in nature, arising in both high energy particle and condensed matter physics. Experimental examples of such systems are the quark gluon plasma state generated in heavy ion collisions~\cite{Adams:2005dq,Aamodt:2010pa}, and quantum quenches of condensed matter systems~\cite{Kibble, Zurek}, for a review and references see e.g.~\cite{QuenchReview}.
The past few years have seen an increased interest in systems far from equilibrium. This renewed interest was spawned by the development of methods allowing the study of such systems at strong coupling~\cite{Chesler:2010bi,Chesler:2008hg,Chesler:2009cy,Chesler:2013lia} utilizing the gauge/gravity correspondence~\cite{Maldacena:1997re}. This correspondence translates the problem of a strongly coupled system far from equilibrium into the equivalent gravitational problem of finding the metric of a time dependent geometry. This task, in general, involves solving the Einstein equations as a system of partial differential equations with a particular set of initial conditions. The main question to be answered within this framework is: what exactly is the full time dependent process of equilibration?

At late times, the question ``How does the system approach equilibrium?'' is identical to the question ``How does the system respond to small perturbations around equilibrium?''.
Depending on the symmetries and details of the gravitational setup, and on the symmetries of the initial conditions, the aforementioned far from equilibrium systems generically asymptote to a particular equilibrium state at late times. In many cases this equilibrium state is going to be some kind of black brane geometry~\cite{Chesler:2013lia,Bhattacharyya:2008jc}. In other words, at late times the fully dynamic nonlinear problem turns into a linear problem of perturbations around black branes. Therefore, one can study the late-time behavior of far from equilibrium systems by considering perturbations around particular black branes, i.e. by computing the quasinormal modes of those black branes. Hence, comparing far from equilibrium results at late times to quasinormal mode results serves as a powerful consistency check of these highly non-trivial calculations.

Quasinormal modes (QNMs) are the ``eigenmodes" of a black brane geometry. Classical black brane horizons are capable of absorbing energy, but can not emit energy. In this sense perturbations around a black brane can lose energy into the horizon, and are thereby a non-conservative system described by a non-self-adjoint operator.
Normal modes of a self-adjoint operator have real-valued frequencies, whereas quasinormal modes associated with a non-self-adjoint operator are complex-valued. Their imaginary part encodes the dissipation of a mode oscillating with the frequency given by their real part. Using the gauge/gravity correspondence the quasinormal modes have been identified with the poles of retarded propagators of the operator dual to the relevant gravity perturbation~\cite{Kovtun:2005ev}. For example, the shear perturbation $h_{xy}$ of the metric is dual to the $xy$-component of the energy momentum tensor $T_{xy}$ within the dual field theory. Therefore, the location of the poles of the retarded propagator $\langle T_{xy} (x_1) \, T_{xy} (x_2) \rangle$ are identical to the quasinormal mode frequencies of the dual gravity perturbation $h_{xy}$.

Numerous results are known for quasinormal modes of black holes, see for example~\cite{Berti:2009kk}, but there are some important gaps in the literature. Hence, in the present paper, we study quasinormal modes of metric (and gauge field) perturbations around two distinct black brane systems: the Reissner-Nordstr\"om black branes and magnetic black branes\footnote{Note that we distinguish between black holes with a compact horizon and black branes which have a planar horizon. In the present paper, however, we are exclusively concerned with black branes.}. 
By magnetic black branes we refer to the solutions of Einstein-Maxwell-Chern-Simons theory in the presence of a constant magnetic field found and analyzed in~\cite{D'Hoker:2009mm,D'Hoker:2009bc,D'Hoker:2010rz,D'Hoker:2010ij}. These magnetic black branes are dual to a $(3+1)$-dimensional field theory at nonzero temperature and with a nonzero magnetic field.
For example, the analysis in the present paper applies to $\mathcal{N}=4$ SYM in a magnetic field for a $U(1)$-subgroup of the R-charge.
In~\cite{D'Hoker:2009mm,D'Hoker:2009bc,D'Hoker:2010rz,D'Hoker:2010ij} such setups were studied with an eye on applications to condensed matter systems in magnetic fields on one hand, and applications to heavy ion collisions on the other~\cite{Fuini:2015hba}. Quasinormal modes of this system have not been computed directly before.

In contrast to that, the Reissner-Nordstr\"om (RN) black brane solution in Anti-de Sitter space (AdS), corresponding to $\mathcal{N}=4$ SYM with a chemical potential and charge density for the R-charge, are well studied. However, while many QNMs have been computed for the $AdS_4$ black holes and black branes~\cite{Berti:2009kk,Konoplya:2011,BertiKokkotas:2003}, the $AdS_5$ Reissner-Nordstr\"om black brane QNMs are not as readily available. Stability of $AdS_5$ RN black holes against metric and electromagnetic perturbations was shown in~\cite{Kodama:2003kk} and~\cite{Konoplya:2008rq}, establishing linearized master equations for perturbations with vanishing spatial momentum. RN black branes are invariant under $SO(3)$ spatial rotations. Hence, their quasinormal modes (at vanishing momentum) can be classified into scalar, vector, and tensor modes because the corresponding perturbations transform as a scalar, vector, or tensor under $SO(3)$ spatial rotations. Some of the $AdS_5$ RN black hole and black brane QNMs have been calculated before~\cite{Berti:2009kk,Konoplya:2011}, in particular the RN black hole scalar QNMs have been studied in~\cite{Konoplya:2008rq} using the formalism developed in~\cite{Kodama:2003jz}, while the RN black brane vector QNMs were studied in~\cite{Maeda:2006by}, and the asymptotic QNMs (those which are infinitely damped) were discussed in~\cite{Natario:2004jd}. 
While~\cite{Maeda:2006by} studies vector QNMs exclusively, \cite{Mas:2006dy} considered a 5-dimensional (single charge) STU black hole and finds an analytic solution for the tensor fluctuation in the hydrodynamic limit, where frequencies and momenta of the perturbation are small compared to the temperature of the black brane. Similarly, in~\cite{Son:2006em} such an analytic solution is found for the tensor fluctuation of the single R-charged RN $AdS_5$ black brane, and in addition the lowest hydrodynamic vector QNM is computed. None of the tensor QNMs are within reach of the hydrodynamic approximation in this case.
Furthermore, within the hydrodynamic limit, also the lowest lying scalar~\cite{Matsuo:2009yu}, vector and tensor~\cite{Ge:2008ak} QNMs have been studied. A Chern-Simons term has been added to the Einstein-Maxwell action in~\cite{Matsuo:2009xn} and influences the fluctuation equations and hence the transport effects, in particular exhibiting chiral transport in the vector fluctuations.
However, the non-hydrodynamic tensor and scalar QNMs of the RN $AdS_5$ black brane have not been computed, as far as we know.
There seems to be no systematic study of the metric tensor fluctuation modes at sizable momentum i.e. beyond the hydrodynamic limit. Neither are we aware of a systematic QNM study of the Reissner-Nordstr\"om black branes over the whole range of charge densities, which runs from zero to the extremal value, $q_{ex}$, at which the black brane has vanishing temperature. In this work we provide a study of the metric tensor fluctuation quasinormal modes at sizable momentum and for a large range of charge densities from zero to over 90 percent of the extremal value.

Two Mathematica~\cite{Wolfram} notebooks, see \cite{notebookLink} for download and details, accompany this publication, allowing the reader to compute quasinormal modes --with any desired accuracy-- for any given charge density and momentum, or alternatively to look up magnetic quasinormal modes for various magnetic field (within reasonable bounds, imposed by numerical accuracy). As a main result, we find that the tensor QNMs of $AdS_5$ Reissner-Nordstr\"om black branes can be divided into two types, referred to as ``Set 1'' and ``Set 2''. Figure~\ref{fig:QNMSets} provides a schematic sketch of these two types of QNMs. Near the extremal charge, we find a set of purely imaginary modes, Set 2, which dominate the late-time behavior at larger charge densities. At small charge densities, these QNMs are too far down the imaginary axis to be seen.\footnote{This fact is in stark contrast to the otherwise similar situation on $AdS_4$ black branes where purely imaginary modes of electromagnetic fluctuations are present near zero frequency already for vanishing charge~\cite{Miranda:2008vb}.}  
Note that also~\cite{Maeda:2006by} show purely imaginary non-hydrodynamic QNMs in their figure 1. However, these modes appear in the vector fluctuations (not in the tensor fluctuations) and their behavior is not discussed in that work.
Following these tensor QNMs of Set 1 while increasing the charge density, we discover kinks in those trajectories, see e.g. figure~\ref{fig:QNMsRN_changeq_k0}. For the magnetic black brane case, we provide the values of the first two tensor QNMs, and the first two scalar QNMs. We find good agreement between our QNMs and the late-time behavior of the far from equilibrium system studied in~\cite{Fuini:2015hba}. These points are discussed in detail in Section~\ref{sec:results}.

%%%%%%%%%%%%%%%%%%%%%%%%%%%%%%%%%%%%%%%
\section{Far from equilibrium setups \& quasinormal modes} \label{sec:FFEnQNM}

The main purpose of the present paper is to provide a benchmark to compare late-time behavior of far from equilibrium setups to. There are in general two steps to this comparison. First, it needs to be determined which equilibrium state is going to be the end point of the equilibration process. This involves specification of parameters such as charge density and magnetic field, as well as matching of the thermodynamics of the far from equilibrium system at late time to the thermodynamics of the equilibrium system.\footnote{For example, one needs to check if both systems are in the same phase, as characterized by the thermodynamic quantities.} Second, at this point in parameter space, the solution of the far from equilibrium system has to be examined as a function of time.\footnote{Note that the same time coordinate has to be chosen in both the far from equilibrium setup as well as the near-equilibrium setup for the comparison of frequencies to be meaningful.} In general, at late times, this solution will be oscillating with a particular frequency and it will simultaneously decay. From this behavior the complex frequency of the most relevant, i.e. the lowest QNM, can be extracted.\footnote{Higher QNMs are also accessible after subtracting the behavior stemming from the lower ones.} This kind of comparison has been successfully employed, for example, in~\cite{Bhaseen:2012gg,Buchel:2015saa}, and~\cite{Fuini:2015hba}.

Depending on the way in which the system is manipulated initially, it will evolve differently. This is reflected in the late-time behavior by which kinds of QNMs describe the approach to equilibrium. Take, for example~\cite{Fuini:2015hba}, i.e. a non-equilibrium system which is initially sheared, between the $xy$-plane and the $z$-direction. This introduces a pressure anisotropy $\Delta p=\langle T^{xx}\rangle-\langle T^{zz}\rangle = \langle T^{yy}\rangle-\langle T^{zz}\rangle$ in the field theory, which corresponds holographically to the metric shear fluctuations $(h_{xx}-h_{zz})$ and $(h_{yy}-h_{zz})$. Here we need to distinguish two cases: 
\begin{itemize}
\item[(i)] The RN black brane solution enjoys invariance under $SO(3)$ spatial rotations when probed with perturbations of vanishing spatial momentum. Then $(h_{xx}-h_{zz})$ and $(h_{yy}-h_{zz})$ are both spin 2 tensors under these $SO(3)$ rotations, and they obey the same equation of motion as the spin 2 tensor fluctuation $h_{xy}$.

\item[(ii)] The magnetic brane solutions are only invariant under $SO(2)$ spatial rotations in the $xy$-plane, because the magnetic field $F_{xy}$ breaks the $SO(3)$ symmetry. In this case $(h_{xx}-h_{zz})$ and $(h_{yy}-h_{zz})$ are both scalars under $SO(2)$ rotations (even at vanishing momentum).
\end{itemize}

This initial shear $\Delta p = \langle T^{xx}\rangle-\langle T^{zz}\rangle$ is the kind of initial condition we have in mind for the main part of this paper. Other initial conditions will require different QNMs at late times. In general, one can identify the required QNMs by the symmetries which are broken by the initial perturbations and the background solution.\footnote{An additional motivation for considering the spin 2 perturbation is the fact that it is related to the shear viscosity through a Kubo formula~\cite{Policastro:2001yc,kss,Kovtun:2004de}.} 
Had the initial condition broken a translational symmetry, for example by a gravitational shock wave initial condition~\cite{Chesler:2010bi,Chesler:2013lia}, more QNMs would be excited. Then it would be interesting to consider more general combinations of perturbations and hence those other QNMs. We stress here, that the relevant fields exhibiting QNMs can be classified according to the symmetry groups of the final equilibration state (e.g. scalars, vectors, tensors under an $SO(3)$ rotation group).

%%%%%%%%%%%%%%%%%%%%%%%%%%%%%%%%%%%%%%%
\section{Holographic setup} \label{sec:setup}

In this section we describe a gravitational system which is dual to a particular strongly coupled plasma in equilibrium, namely $\mathcal{N}=4$ Super-Yang-Mills theory at nonzero temperature. Two distinct setups are discussed, one corresponding to a charged plasma, the second corresponding to a neutral plasma within an external magnetic field. Both of these brane configurations are solutions to the five-dimensional Einstein-Maxwell equations. 
They are derived from the action
\begin{equation}\label{eq:EMAction}
S = \frac{1}{2\kappa} \int d^5 x \, \sqrt{-g} \left [
(R-2\Lambda) - L^2 F_{\mu\nu} F^{\mu\nu}
\right ] \, ,
\end{equation}
with $\kappa = 8\pi G_N$ for the five-dimensional Newton constant $G_N$. A Chern-Simons term may be added to this action, but would influence neither of the black brane solutions we are interested in.\footnote{However, such a term potentially changes the spin 0 (scalar) and spin 1 (vector) fluctuation equations.}

%--------------------------------------------------------
\subsection{Reissner-Nordstr\"om black branes}
Here we discuss the Anti-de Sitter Reissner-Nordstr\"om black brane solution to \eqref{eq:EMAction}, which is dual to a charged $\mathcal{N}=4$ Super-Yang-Mills plasma at large $N$ and large 't Hooft coupling $\lambda$.

%.................................
\subsubsection{Equilibrium solutions and thermodynamics} \label{sec:RNbranesEquilibrium}

The Reissner-Nordstr\"om black brane metric in the Poincare patch is defined by
\begin{equation} \label{eq:rRNMetric}
ds^2 = \frac{r^2}{L^2}\left (
-f dt^2 +{d\vec x}^2 
\right )+\frac{L^2}{r^2 f} dr^2  \, , 
\end{equation}
with the blackening factor given by
\begin{equation} \label{eq:blackeningFactor}
f(r) = 1 - \frac{m L^2}{r^4} + \frac{q^2 L^2}{r^6} \, .
\end{equation}
This geometry has a boundary at $r=\infty$ and an outer (non-compact) horizon at some $r=r_H\ge 0$. The blackening factor \eqref{eq:blackeningFactor} has six roots of which we identify the largest positive one with $r_H$. At the extremal charge value $q=q_{ex}$ two of the real roots coincide. As we will see later, this implies that the fluctuation equations of motion have coefficients with irregular singular points, while these coefficients have only regular singular points for all other $q<q_{ex}$. The charge of the black brane couples to a $U(1)$-gauge field which takes the form
\begin{equation}
A_t = \mu - \frac{Q}{L \, r^2} \, ,
\end{equation}
where $Q=\frac{\sqrt{3}q}{2}$ for this to be a solution to the Einstein-Maxwell equations. 

In the rest of this paper we are going to measure quantities in units of $L$, which amounts to setting $L=1$. Note that this background with $m=1$ is identical to the background chosen in \cite{Policastro:2002se} after setting the charges to zero and performing the coordinate transformation to the coordinate $u=r_H^2/r^2$.

Let us collect the expressions for the relevant thermodynamic quantities in the dual field theory. Requiring regularity of the Euclideanized manifold at the horizon, the temperature and chemical potential are
\begin{eqnarray}
T &=& r_H^2\frac{|f'(r_H)|}{4\pi} \, ,\\
\mu &=& \frac{\sqrt{3} q}{2 r_H^2} \, .
\end{eqnarray}
The extremal charge is determined to be $q_{ex} = \sqrt{2} \left (\frac{m}{3}\right )^{3/4}$, derived from the condition that $T=0$ for the extremal Reissner-Nordstr\"om black brane.

It will be useful to transform and rescale our metric for the computation of the quasinormal frequencies. We start from \eqref{eq:rRNMetric} and perform the transformation $r^2\to r_H^2/u$. In these coordinates, the $AdS$ boundary is located at $u=0$, while the outer horizon is located at $u=1$. We further rescale $q\to \tilde q\, r_H^{3}$. After these transformations and rescalings we get the metric
\begin{equation} \label{eq:zRNMetric}
ds^2 = \frac{r_H^2}{u}\left (
-f(u) dt^2 +{d\vec x}^2 
\right )+\frac{du^2}{4 u^2 f(u)}   \, , 
\end{equation}
with $f(u) = 1 - m \, r_H^{-4} u^2 + {\tilde q}^2 u^3$, and we may use the horizon condition $0=f(u=1) \Rightarrow m \, r_H^{-4} = 1+{\tilde q}^2$ in order to remove the dependence on $m$. Then the metric depends only on two parameters namely $r_H$ and $\tilde q$, while the blackening factor reduces to $f(u) = 1 - (1+{\tilde q}^2) u^2 + {\tilde q}^2 u^3$. \footnote{\label{ft:rhdep}The dependence on $r_H$ can be eliminated by a rescaling of $t$ and $\vec{x}$. Of course, we could alternatively remove all dependence on $\tilde q$ in favor of dependence on $m$.}

We also re-write the thermodynamic quantities in these coordinates:
\begin{eqnarray}
T &=& \frac{|f'(u=1)|}{2 \pi} r_H = \frac{2 - {\tilde q}^2}{2 \pi} r_H
=\frac{1}{\pi}(1- \frac{\tilde{q}^2}{2}) \left (\frac{m}{1+\tilde{q}^2}\right)^{\frac{1}{4}}
 \, ,\\
\tilde \mu &=& \frac{\sqrt{3} \tilde q}{2} r_H  \, ,
\end{eqnarray}
and the extremal charge is given by $\tilde q_{ex} = \sqrt{2}$. Note that the relation between the original charge $q$ and our rescaled $\tilde q$ in these coordinates can be expressed in a closed form
\begin{equation} \label{eq:qtOfq}
q = \tilde q \, r_H^{3} = 
\tilde q \left ( \frac{m}{1+{\tilde q}^2} \right )^{3/4} \, .
\end{equation}
The entropy density is given by the horizon area, and it reads
\begin{equation}
s_{\text{RN}} = \frac{\sqrt{g_{(3)}}|_{r_H}}{4\pi} = \frac{r_H^{3}}{4\pi} \, ,
\end{equation}
where $g_{(3)}$ is the determinant of the spatial metric induced on the horizon.

The energy density and pressure of the Reissner-Nordstr\"om black brane can be computed using the standard transformation to Fefferman-Graham coordinates following~\cite{deHaro:2000xn, Skenderis:2002wp}. This procedure yields
\begin{eqnarray}
\epsilon_{\text{RN}} & = & \frac{3}{2\kappa} m \, , \\
P_{\text{RN}} & = & \frac{1}{2\kappa} m \, , 
\end{eqnarray}
where $\kappa$ was defined in and below equation \eqref{eq:EMAction}.
Note that this implies tracelessness for the energy momentum tensor
\begin{equation}
\langle T^{\mu\nu} \rangle = 
\frac{1}{2\kappa} \left (
\begin{array}{c c c c}
{3} m & 0 & 0 & 0 \\
0 & m & 0 & 0 \\
0 & 0 & m & 0 \\
0 & 0 & 0 &  m 
\end{array}
\right) \, ,
\end{equation}
such that $\langle T_\mu{}^\mu\rangle = 0$. Even though, our ground state has nonzero temperature and chemical potential, both breaking the conformal symmetry, the action of $\mathcal{N}=4$ Super-Yang-Mills theory is still invariant under conformal transformations and hence operator identities are valid.\footnote{Because operator identities are independent of the state.} 
In particular, conformal symmetry implies tracelessness of the energy-momentum tensor up to contributions from the conformal anomaly given by $T^\mu_\mu = - a E_4 - c I_4 -\alpha \, F^2$, with the anomaly coefficients $a,\, c,\, \alpha$. Here, $E_4$ and $I_4$ contain various contractions of the Riemann curvature tensor, which vanish as our field theory lives in flat spacetime. The third term $-\alpha\, F^2$ accounts for the presence of an external field strength $F_{\mu\nu}$, which also vanishes in the ground state discussed in this section.

%.................................
\subsubsection{Fluctuations} \label{sec:RNfluc}

Having worked out the RN $AdS_5$ black brane metric $g_{\mu\nu}$, we are now ready to introduce fluctuations $\delta g_{\mu\nu}$ around this background. In this section we collect the linearized Einstein equations obeyed by these metric fluctuations. We choose to write these equations in momentum space, after the Fourier transformation $\delta g_{\mu\nu} \equiv e^{-i\omega t + i k z} \, h_{\mu\nu}(\omega, k,u)$. Throughout this paper, we choose the radial gauge $h_{r \mu} \equiv 0$.

Metric shear fluctuations are 2-tensors under the rotation group, and thus decouple from all other fluctuations. Shear fluctuations, such as $h_{xy}$, satisfy an equation of motion which --after a field redefinition $\phi=h_x^y = g^{yy} h_{xy}$-- can be written as
\begin{eqnarray} \label{eq:hxy}
0 & = &\phi'' - \frac{f(u)-u\,f'(u)}{u \,f(u)} \phi' +\frac{\omega^2- f(u) k^2}{4 r_H^2 u\,f(u)^2} {\phi}
\, .
\end{eqnarray}
We will often work with the rescaled quantities $\tilde{\omega}\equiv \omega/r_H$ and $\tilde{k}\equiv k/r_H$ as this removes all factors of $r_H$ from the equation of motion. This is equivalent to the coordinate rescaling mentioned in footnote \ref{ft:rhdep}. Note that, with the momentum in the $z$-direction, there exists one more tensor excitation given by $(h_{xx}-h_{yy})$. This combination also satisfies the tensor equation \eqref{eq:hxy} under the replacement $h_{xy} \to (h_{xx}-h_{yy})$. Setting the momentum to zero allows two additional tensor fluctuations, namely $(h_{xx}-h_{zz})$ and $(h_{yy}-h_{zz})$. We have checked that in this case all four of these tensor fluctuations satisfy equation \eqref{eq:hxy} with $k\equiv0$, and they again decouple from all other fluctuation equations. In the present paper, we are interested in these sets of $SO(2)$ tensor fluctuations $h_{xy},\, (h_{xx}-h_{yy})$ (for nonzero $k$), and $SO(3)$ tensor fluctuations $h_{xy}, \, (h_{xx}-h_{yy}),\, (h_{xx}-h_{zz}),\, (h_{yy}-h_{zz})$ (for $k=0$). As mentioned in the introduction, these tensor fluctuations correspond to pressure anisotropies $\Delta p = (\langle T^{xx}\rangle - \langle T^{zz}\rangle)$ near equilibrium. Due to the rotational symmetry in the $xy$-plane, we know $\langle T^{xx}\rangle=\langle T^{yy}\rangle$. At late times, a system which was initially sheared between the $xy$-plane and the $z$ direction, should be dominated by fluctuations $(h_{xx}-h_{zz}),\, (h_{yy}-h_{zz})$ at vanishing momentum. For completeness we discuss here also the linearized Einstein equations obeyed by vector and scalar fluctuations.

Vector fluctuations satisfy the two coupled equations
\begin{eqnarray}\label{eq:hxi}
0 & = & 
{h_{ti}}''+\frac{{h_{ti}}'}{u}-\frac{{h_{ti}}}{u^2}-2\sqrt{3} {\tilde q}  {a_i}'
 \, ,
\end{eqnarray}
and 
\begin{equation}\label{eq:maxwelli}
0 =
{a_i}''+\frac{u \left(3 {\tilde q}^2 u-2 m\right)}{f(u)} {a_i}'+\frac{\tilde{\omega}^2}{4 u \,f(u)^2}{a_i}-\frac{\sqrt{3} {\tilde q} \left(u {h_{ti}}'+h_{ti}\right)}{2 f(u)}
 \, ,
\end{equation}
with the gauge field fluctuations $a_\mu$, while indices take the values 
$i=1,2,3$ at $k=0$.
Using the constraint equation for $h_{ti},\, a_i$ resulting from the $ri$-component of the Einstein equation, this can be transformed to a single equation
\begin{equation}
0 = {a_i}''
+\left[ \frac{1}{-1 + u} + \frac{1 - 2 \tilde q^2 u}{1 + u - \tilde q^2 u^2}\right ] {a_i}'
+\frac{-12 \tilde q^2 (-1 + u) u^2 (-1 + u (-1 + \tilde q^2 u)) + \tilde\omega^2}{4 (-1 + 
   u)^2 u (1 + u - \tilde q^2 u^2)^2} {a_i} \, .
\end{equation}

Together, the vector and tensor fluctuations contain all the quasinormal modes of the Reissner-Nordstr\"om black brane. Let us consider the case $k=0$ first. There are coupled equations for the scalar fluctuations $h_{tt},\, (h_{xx}+h_{yy}),\, a_t$. But adding and subtracting those equations adequately to/from each other shows that neither of these posess any non-trivial quasinormal mode frequencies. In other words, these scalar fluctuations are completely determined by boundary data. In contrast to that, switching on momentum in the $z$ direction, $k\neq 0$, the scalar fluctuations which couple to each other are given by $h_{tt},\, (h_{xx}+h_{yy}+h_{zz}), \, h_{tz},\, a_t, \, a_z$. This set of scalar fluctuations has non-trivial QNMs, for example sound modes with a dispersion $\omega \propto v_s \, k$ for a speed of sound $v_s$. We are not concerned with these latter QNMs in this paper and focus instead on the shear fluctuations.

In order to find a solution to equation \eqref{eq:hxy}, we need to specify two boundary conditions. At the horizon we find the solution behaves as
\begin{equation}\label{eq:nearHorizonhxy}
h_{xy} = (1-u)^{\pm \frac{i\tilde{\omega}}{2 (2-{\tilde q}^2)}} \left [
h^{(0)} + h^{(1)} (1-u) + ...
\right ] \, ,
\end{equation}
where the negative (positive) sign corresponds to the infalling (outgoing) solution. Analogously, we have \begin{equation}\label{eq:nearHorizonax}
a_{i} = (1-u)^{\pm \frac{i\tilde{\omega}}{2 (2-{\tilde q}^2)}} \left [
a^{(0)} + a^{(1)} (1-u) + ...
\right ] \, ,
\end{equation}
for the vector modes.
Quasinormal modes are defined as those solutions to the linearized Einstein equations which: 
\begin{itemize}
\item obey the infalling boundary condition at the black brane horizon, corresponding to the negative sign in equation \eqref{eq:nearHorizonhxy} and \eqref{eq:nearHorizonax}, and
\item satisfy a Dirichlet condition at the boundary of $AdS$ space.
\end{itemize}
Since these two boundary conditions are imposed at two distinct points, we need to work out a method to find solutions obeying both conditions simultaneously. This will be our goal in Section \ref{sec:methods}.  

%--------------------------------------------------------
\subsection{Magnetic black branes} \label{sec:magBranes}

In contrast to the previous section, let us now consider a setup with vanishing electric charge, but in the presence of a constant magnetic field $F_{xy}$. One such solution of action \eqref{eq:EMAction} is the magnetic black brane~\cite{D'Hoker:2009mm,D'Hoker:2009bc,D'Hoker:2010rz,D'Hoker:2010ij} which is dual to an uncharged $\mathcal{N}=4$ Super-Yang-Mills plasma in an external magnetic field at large $N$ and large 't Hooft coupling $\lambda$.

%.................................
\subsubsection{Equilibrium solutions and thermodynamics}

For the metric and field strength we choose the following ansatz~\cite{D'Hoker:2009mm}
\begin{eqnarray}
ds^2 & = & -U(r) dt^2 + \frac{dr^2}{U(r)}+e^{2 V(r)} ({dx}^2+{dy}^2) +e^{2 W(r)} dz^2  \, ,\\ 
F & = & b dx \wedge dy   \, .
\end{eqnarray}
Then we transform to the $u$ coordinates via $r=r_H u^{-1/2}$, and rescale $U\to r_H^2 \tilde U$ yielding
\begin{eqnarray} \label{eq:magBraneMetric}
ds^2 & = & -r_H^2 \tilde U(u) dt^2 + \frac{du^2}{4 u^3 \tilde U(u)}+e^{2 V(u)} ({dx}^2+{dy}^2) +e^{2 W(u)} dz^2  \, ,\\ 
F & = & b dx \wedge dy   \, .
\end{eqnarray}
We are interested in asymptotically $AdS_5$ solutions to the equations of motion following from the action \eqref{eq:EMAction}, hence we set the cosmological constant $\Lambda$ to its $AdS_5$ value, namely $\Lambda=-6$ in units of the $AdS$ radius $L$.
This casts the Einstein-Maxwell equations into the form
\begin{eqnarray}
0 &=& 
2 b^2+4 e^{4 V(u)} \big(u^3 \tilde U'(u) (2 V'(u)+W'(u))
+u^2 \tilde U(u) (2 (u (2 V''(u) \nonumber\\&&
+W''(u)+W'(u)^2)
+V'(u) (2 u W'(u)+3)+3 u V'(u)^2)+3 W'(u))-3\big)
\, , \nonumber \\ \nonumber
0 &=& 2 u^2 e^{4 V(u)} (2 u \tilde U''(u)+\tilde U'(u) (4 u (V'(u)+W'(u))+3)
+\tilde U(u) (4 u (V''(u) \\&&
+W''(u)+W'(u)^2)
+V'(u) (4 u W'(u)+6)+4 u V'(u)^2+6 W'(u)))-2 (b^2+6 e^{4 V(u)}) \, , \nonumber \\ \nonumber
0 &=& 
b^2 e^{-4 V(u)}+u^2 (2 (u \tilde U''(u)\\&&
+\tilde U(u) (4 u V''(u)+6 V'(u) (u V'(u)+1)))+\tilde U'(u) (8 u V'(u)+3))-6
\, , \nonumber \\
0 &=& 
b^2 e^{-4 V(u)}+2 u^3 (\tilde U'(u) (2 V'(u)+W'(u))+2 \tilde U(u) V'(u) (V'(u)+2 W'(u)))-6
 \, . \label{eq:UVW}
\end{eqnarray}
These are the equations of motion which we solve numerically in order to determine $\tilde U, \, V,\, W$ as functions of $b$ and of the temperature, which is associated with the horizon value of $\tilde U'(u)$. 
All of our calculations will be performed in this coordinate system since it is convenient for the background. If needed, we will rescale quantities afterward in order to obtain physical values.

We solve the equations of motion with the following boundary conditions at the horizon $u=1$:
\begin{eqnarray} \label{eq:horizonVW}
\tilde U &=& u_0 + u_1 (1-u)+ u_2 (1-u)^2 + \mathcal{O}((1-u)^2) \, , \\
V &=& v_0 + v_1 (1-u) + \mathcal{O}((1-u)^2) \, , \\
W &=& w_0 + w_1 (1-u) + \mathcal{O}((1-u)^2) \, ,
\end{eqnarray}
where $u_0=0$ (in order to obtain a spacetime with horizon) and we can set $v_0=w_0=0$ by a rescaling of the coordinates $x$, $y$, and $z$. The equations of motion imply
\begin{eqnarray}
u_2&=& \frac{-6+5 b^2 e^{-4 v_0}+9u_1}{12} =   \frac{-6+5 b^2+9u_1}{12}\, , \\
v_1 &=&\frac{3-b^2 e^{-4 v_0}}{3 u_1} = \frac{3-b^2}{3 u_1} \, ,\\
w_1&=& \frac{b^2 e^{-4 v_0}+6}{6 u_1} =\frac{b^2 +6}{6 u_1} \, .
\end{eqnarray}
Note that we in fact retain ten orders in the near horizon expansion in order to obtain better numerical solutions. However, we refrain from reproducing the lengthy expressions (for those $u_3,\, u_4,\, ...$, as well as $v_2,\, v_3,\, ...$, and $w_2,\, w_3, ...$) here. From the near-horizon expansion above we see that the temperature is associated with the choice of the parameter $u_1$. The magnetic field will be associated with the choice of the parameter $b$. Choosing $b$ and $u_1$ completely determines the solution to our equations of motion.

Demanding an asymptotically AdS solution, these functions near the boundary $u=0$ behave like
\begin{eqnarray} \label{eq:boundaryExpUVWshort}
\tilde U = \frac{1}{u} + \frac{u^B_{(1)}}{\sqrt{u}}+ \mathcal{O}(u^0, u \log u) \, , \nonumber\\
e^{2 V} = \frac{v}{u} +\frac{v u^B_{(1)}}{\sqrt{u}}+ \mathcal{O}(u^0, u \log u) \, , \\ \nonumber
e^{2 W} = \frac{w}{u} +\frac{w u^B_{(1)}}{\sqrt{u}}+ \mathcal{O}(u^0, u \log u) \, ,
\end{eqnarray}
where $v$ and $w$ are dimensionless parameters depending on the choice of $u_1$ and $b$. Here, $u^B_{(1)}$ is the subleading coefficient in the near boundary expansion of $\tilde{U}$, which will be discussed below equation~\eqref{eq:UVWBdy}. 
We would like to rescale the coordinates such that the metric asymptotes to canonical $AdS_5$ with a Minkowskian boundary. 
Hence, we rescale $t\to \hat t = r_H\,t$, $x \to  \hat x = \sqrt{v} x$, $y \to  \hat y = \sqrt{v} y$, $z \to \hat z = \sqrt{w} z$, which near the boundary gives
\begin{eqnarray} \label{eq:hattedCoordMetric}
ds^2 & \approx & \left [-\frac{1}{u} d{\hat t}^2+\frac{1}{u} ({d{\hat x}}^2+{d{\hat y}}^2) + \frac{1}{u} d{\hat z}^2\right ]  + \frac{du^2}{4 u^2} \, ,\\ 
F & \approx & \frac{b}{v} d{\hat x} \wedge d{\hat y}   \, .
\end{eqnarray}
Since none of our rescalings involves the radial coordinate $u$, the functions $\tilde U(u),\, V(u),\, W(u)$ remain unchanged.
The full bulk metric after these rescalings reads
\begin{eqnarray} \label{eq:gHat}
ds^2 & = & -{\tilde U} d{\hat t}^2 + \frac{du^2}{4 u^3 \tilde U}+\frac{e^{2V}}{v} ({d{\hat x}}^2+{d{\hat y}}^2) + \frac{e^{2W}}{w} d{\hat z}^2 \, ,\\  \label{eq:FHat}
F & = & \frac{b}{v} d{\hat x} \wedge d{\hat y}   \, .
\end{eqnarray}
In summary, in these hatted coordinates the $AdS$ defining function~\cite{MR837196} can be taken to simply be $u$, giving a Minkowskian boundary metric.

The physical magnetic field, as defined through equation~\eqref{eq:FHat}, and the temperature in the boundary field theory are given by 
\begin{equation}
\mathcal{B} = \frac{b}{v}\, \qquad T=\frac{u_1}{2\pi }\, .
\end{equation}
The entropy density is determined by the area of the horizon
\begin{equation}
s = \left . \frac{1}{4\pi \, v\sqrt{w}} e^{2V+W}  \right |_{r_H} =  \frac{1}{4\pi \, v\sqrt{w}} \, ,
\end{equation}
where near the horizon $e^{2V+W}\to 1$ as can be seen from the expansion \eqref{eq:horizonVW} with $v_0=0=w_0$.

In order to obtain the energy density and pressure of the dual field theory we need to work a little harder. We again follow~\cite{deHaro:2000xn, Skenderis:2002wp} in order to extract the energy momentum tensor from the near-boundary metric in Fefferman-Graham coordinates \cite{MR837196}. First, we need the boundary expansion for the background fields in our present coordinates $\hat t,\, \hat x,\, \hat y,\, \hat z,\, u$:
\begin{eqnarray}\label{eq:UVWBdy}\nonumber
\tilde U & = & \frac{1}{u} +\frac{u^B_{(1)}}{\sqrt{u}} + \frac{{u^B_{(1)}}^2}{4} + u^B_{(4)} u + u \log u \frac{b^2}{3 v^2} +\mathcal{O}(u^{3/2},\, u^{3/2}\log u )  \, ,\\ 
e^{2 V} & = & \frac{v}{u} + {v u^B_{(1)}} \frac{1}{\sqrt{u}} + \frac{v {u^B_{(1)}}^2}{4} + v^B_{(4)} u - u \log u \frac{b^2}{6 v } +\mathcal{O}(u^{3/2},\, u^{3/2}\log u ) \, ,\\ \nonumber
e^{2 W} & = &\frac{w}{u} + {w u^B_{(1)}} \frac{1}{\sqrt{u}} + \frac{w {u^B_{(1)}}^2}{4} -\frac{2 w v^B_{(4)}}{v} u + u \log u \frac{w b^2}{3 v^2 } +\mathcal{O}(u^{3/2},\, u^{3/2}\log u )  \, ,
\end{eqnarray}
where $u^B_{(4)}$ and $v^B_{(4)}$ are free parameters. The parameter $u^B_{(1)}$ is not free, it in fact can be removed by a residual diffeomorphism invariance of the metric ansatz \eqref{eq:gHat}: sending $u\to u/(1-u^B_{(1)}\sqrt{u}/2)^2$ leaves this metric invariant, and will set $u^B_{(1)}\to 0$ in the expansion \eqref{eq:UVWBdy}. In this way $u^B_{(1)}$ is seen to be related to the horizon location.

Next, we derive the coordinate transformation which takes us from the present coordinate system to Fefferman-Graham coordinates which can be defined by having a radial coordinate $r$ such that $g_{rr}\equiv 1/r^2$ throughout the entire spacetime, as well as the radial gauge $g_{r\mu}=0$ as above. The transformation is defined by equating the line element in Fefferman-Graham coordinates
\begin{equation} \label{eq:gFG}
{ds^2}_{\text{FG}} \equiv  \frac{1}{r^2}\left[  
dr^2 + (g^{(0)} + r^2 g^{(2)}+ r^4 g^{(4)} +h^{(4)} r^4 \log r^2 +\mathcal{O}(r^3))_{ij} dz^i dz^j\right]
\, ,
\end{equation} 
to the line element in our current coordinates, i.e.
\begin{equation}\label{eq:ds2Isds2}
{ds^2}_{\text{FG}} = {ds^2}\, ,
\end{equation}
where $ds^2$ is given by equation \eqref{eq:gHat}.
Our ansatz is that the metric component coefficients $g^{(n)}$ and $h^{(n)}$ do not depend on the Fefferman-Graham radial coordinate $r$, that $z_0=\hat t,\, z_{1} = \hat x, \, z_2=\hat y,\, z_3=\hat z$ and $r$ is a function of only the radial $u$ coordinate, i.e. $r=r(u)$.  Hence near the boundary
\begin{equation}\label{eq:rBdy}
r(u) = r^B_{(1)} \sqrt{u} + r^B_{(2)} u +  r^B_{(3)} u^{3/2} + r^B_{(4)} u^2+ r^B_{(5)} u^{5/2} + r^B_{(L,5)} u^{5/2} \log u + \mathcal{O}(u^3, u^3 \log u)\, ,
\end{equation}
where we will choose $r^B_{(1)}=1$ so that the boundary metric remains canonical Minkowskian. Now we plug the boundary expansions for $r(u)$, equation \eqref{eq:rBdy}, the expansions for $\tilde U(u),\, e^{2V(u)},\, e^{2W(u)}$ given by equation \eqref{eq:UVWBdy}, and the metric expansion \eqref{eq:gFG}, into the line element equation \eqref{eq:ds2Isds2}. Expanding the resulting equation around the boundary $u=0$ and determining coefficients order by order we obtain the transformation for the radial coordinate near the boundary
\begin{eqnarray}
r(u) &=& 
\sqrt{u} -\frac{u^B_{(1)}}{2} u + \frac{{u^B_{(1)}}^2}{4} u^{3/2} - \frac{{u^B_{(1)}}^3}{8} u^2 - u^{5/2} \log u \frac{b^2}{24 v^2} + \mathcal{O}(u^3,\,u^3 \log u)
 \, .
\end{eqnarray}
In Fefferman-Graham coordinates the near boundary metric has the expansion 
\begin{eqnarray}
g_{00} &=& -1 -r^4 \frac{b^2 + 18 v^2 u^B_{(4)}}{24 v^2} - r^4 \log r \frac{b^2}{2 v^2}+\mathcal{O}(r^5)\, ,\\
g_{11}= g_{22} &=& 1 +r^4 \frac{b^2 - 6 v^2 u^B_{(4)}+24 v v^B_{(4)}}{24 v^2} - r^4 \log r \frac{b^2}{2 v^2}+\mathcal{O}(r^5)\, ,\\
g_{33} &=& 1 +r^4 \frac{b^2 - 6 v (v u^B_{(4)}+8 v^B_{(4)})}{24 v^2} + r^4 \log r \frac{b^2}{2 v^2}+\mathcal{O}(r^5)\, .
\end{eqnarray}
In this case, the holographic energy-momentum tensor is given by \cite{deHaro:2000xn,Fuini:2015hba}
\begin{eqnarray}\label{eq:Tij}
\langle T_{ij} \rangle \equiv \frac{2}{\kappa}
\left[
g^{(4)}_{ij} - g^{(0)}_{ij}tr \, g^{(4)} -\left(\log(\Lambda)+\mathcal{C}\right)h^{(4)}_{ij} \right]\, , 
\end{eqnarray}
where $\Lambda$ is the renormalization scale which we will pick to be $\Lambda=\sqrt{\mathcal{B}}$, and $\mathcal{C}$ is an arbitrary constant reflecting this ambiguity. We will choose $\mathcal{C}=-1/4$, as this eliminates the $b^2$ contribution to the energy density arising solely from the background magnetic field. The logarithmic contributions, $h^{(4)}$, are associated with the conformal anomaly~\cite{Petkou:1999fv}. From  equation \eqref{eq:Tij} we obtain the energy density and pressures
\begin{eqnarray}\label{eq:epsilonMagBrane}
\epsilon &\equiv& \langle T_{00} \rangle = \frac{2}{\kappa}  \left[-
\frac{3}{4}u^B_{(4)} + \frac{\mathcal{B}^2}{4} \log \mathcal{B} \right] \, , \\
P_1=P_2 &\equiv& \langle T_{11} \rangle = \frac{2}{\kappa}  \left[
-\frac{1}{4}u^B_{(4)}+\frac{v^B_{(4)}}{v}-\frac{\mathcal{B}^2}{4}+\frac{\mathcal{B}^2}{4} \log\mathcal{B}
\right] \, , \\
P_3 &=& \langle T_{33} \rangle =\frac{2}{\kappa}  \left[
-\frac{1}{4}u^B_{(4)}-2\frac{v^B_{(4)}}{v} -\frac{\mathcal{B}^2}{4} \log\mathcal{B}
\right] \, .
\end{eqnarray}
Had we chosen a different renormalization scale $\Lambda = 1$,\footnote{This is really $\Lambda=1/L$ in units of the curvature radius.} this would remove all explicit $\mathcal{B}$-dependence from the energy density, i.e. we would get $\langle T_{00}\rangle_{\Lambda=1} =  -\frac{2}{\kappa} \frac{3}{4}u^B_{(4)}$. 

\begin{figure}[]
\includegraphics[height=7cm]{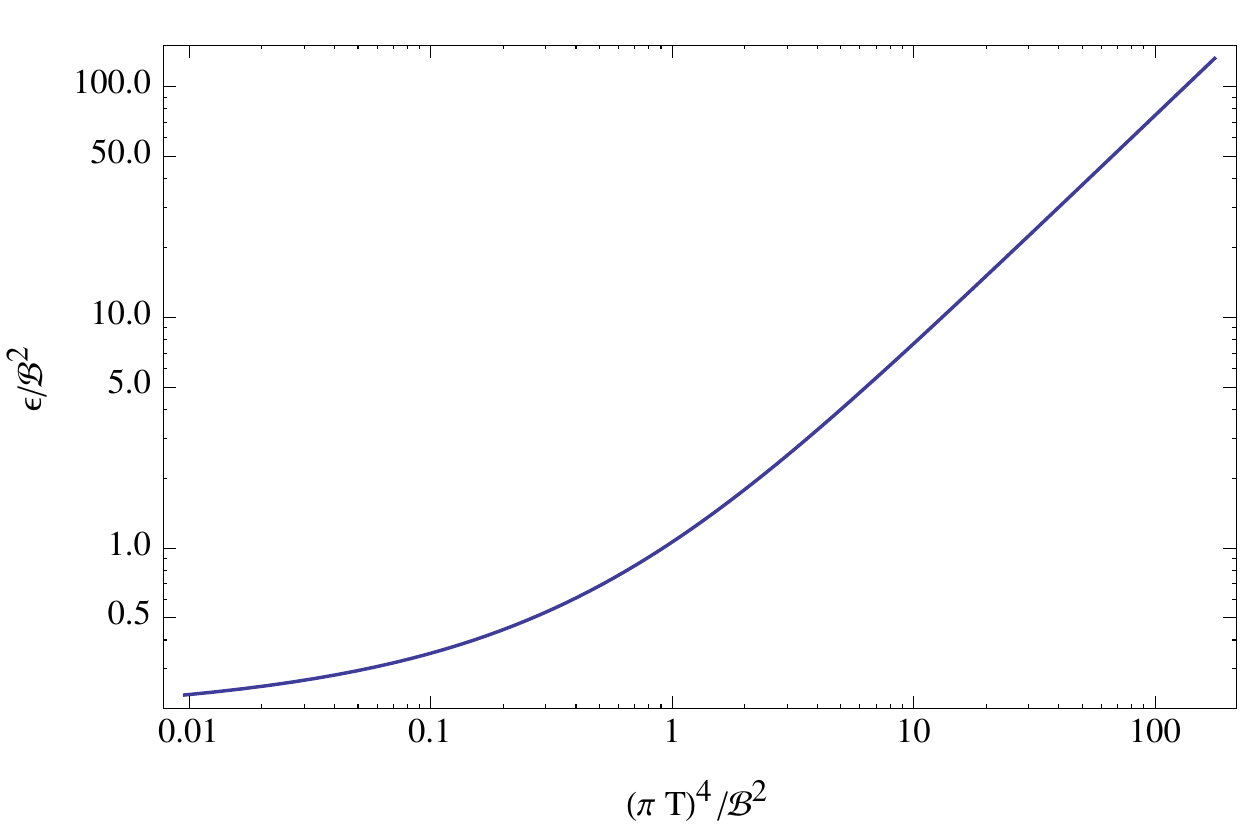}
\caption{\label{fig:epsilonTBmagBrane}
Energy density $\epsilon$ of the field theory dual to the magnetic brane solution as a function of the dimensionless ratio $(\pi T)^4/\mathcal{B}^2 $ between the two scales in the problem, namely the boundary temperature $T$ and boundary magnetic field $\mathcal{B}$. At small magnetic field values, $\mathcal{B}\ll (\pi T)^2$, the energy density approaches the value of the Schwarzschild $AdS_5$ black brane, namely $\epsilon = 3/4 (\pi T)^4$.
}
\end{figure}

The energy density given in \eqref{eq:epsilonMagBrane} apparently depends on both, magnetic field $\mathcal{B}$ and temperature $T$, seemingly independently. However, we know that the system arises from a conformal setting. Thus we expect that dimensionless observables should only depend on the ratio $T^2/\mathcal{B}$ of these two scales, see~\cite{Fuini:2015hba}. Therefore, it suffices to show the energy density \eqref{eq:epsilonMagBrane}, divided by $\mathcal{B}^2$, as a function of $T^4/\mathcal{B}^2$ in Fig.~\ref{fig:epsilonTBmagBrane}. 
Note, that our numerical data satisfies this scaling relation fairly well, as we have checked by changing the two horizon parameters $u_1$ and $b$. However, due to accumulating numerical errors there can be deviations of up to 3 percent in our current data. More precisely, there is an error in our energy density that is on the order of 0.19 \% for intermediate magnetic field ($\mathcal{B}/T^2=0.99$), and 2.3 \% for large magnetic fields ($\mathcal{B}/T^2=29.5$).
This error can be reduced by requiring higher precision. We refrain from pushing these deviations to zero as the current accuracy suffices for our purposes. The far-from equilibrium data we intend to compare to is currently known only up to a comparable error~\cite{Fuini:2015hba}.

Note that the trace of the energy momentum tensor is given by
\begin{equation}\label{eq:traceTij}
 \langle T_i^i \rangle = -\frac{2}{\kappa} 
\frac{b^2}{2 v^2} = -\frac{\mathcal{B}^2}{\kappa} \, .
\end{equation}
This is a manifestation of the trace anomaly due to an external field $F$, as discussed in the last paragraph of Section \ref{sec:RNbranesEquilibrium}. In the case at hand, that external field is simply the physical magnetic field $\mathcal{B}=b/v$.
Note that according to \eqref{eq:traceTij} the energy momentum tensor is traceless in the limit of vanishing magnetic field, as expected.

%----------------------------------------------------------------------------
\subsubsection{Fluctuations} \label{sec:fluctuationsMagnetic}
In this section, we consider two types of fluctuations. First, we discuss the tensor fluctuations $h_{xy}$ and $h_{xx}-h_{yy}$, which are 2-tensors under the rotational $SO(2)$ remaining in presence of the magnetic field $F_{xy}$~\footnote{In this section we work exclusively with hatted coordinates. So, technically, all of the coordinates $t,\, x,\, y,\, z$ and vector/tensor components such as $h_{xy}$ should read $\hat t,\, \hat x,\, \hat y,\, \hat z$ and $h_{\hat x\hat y}$, respectively. In order not to clutter our equations, we drop the hats in this section.}. Second, we discuss the fluctuations $(h_{xx}-h_{zz})$ and $(h_{yy}-h_{zz})$, which are now scalars under the $SO(2)$ and therefore couple to the other scalar perturbations. This set of scalar perturbations gives rise to the QNMs which describe the late-time behavior of the initially sheared systems we are interested in.

Linearizing the Einstein-Maxwell equations in the fluctuations, again picking radial gauge $h_{r \mu} \equiv 0$, we find that the mode $(h_{xx} - h_{yy})$ decouples from all others and satisfies the same equation as the tensor mode $h_{xy}$.
After a Fourier transformation $h_{\mu\nu}(x) \propto e^{-i\hat{\omega} \hat{t} + i \hat{k} \hat{z}} \, h_{\mu\nu}(\hat{\omega}, \hat{k})$, that fluctuation equation is given by
\begin{eqnarray} \label{eq:hxyMagnetic}
0 &=& h_{xy}''+(\frac{\tilde U'}{\tilde U}-2 V'+W'+\frac{3}{2 u}) h_{xy}' \nonumber \\ \nonumber
&&+\frac{1}{4 u^3 \tilde U^2}(\tilde U (-2 b^2 e^{-4 V}+4 u^3 \tilde U'' +2 u^2 \tilde U' (4 u W'+3)-12)\\
&&+4 u^2 \tilde U^2 (2 u (V'^2+W''+W'^2)
+3 W')+\hat{\omega} ^2-\hat{k}^2 \tilde U e^{-2 W})  h_{xy}  \, ,
\end{eqnarray}
which, for $b=0$, correctly reduces to the equation of a minimally coupled scalar in an $AdS_5$ black brane geometry once we redefine $h_{xy}= \phi/ u$, and choose $\tilde U=\frac{1}{u} -u ,\,  V =  W= -\frac{1}{2} \log u$. We note in passing that the equation \eqref{eq:hxyMagnetic} at nonzero magnetic field is not simply that of a minimally coupled scalar field in the magnetic black brane background.

Again, we make the ansatz $h_{xy}=(1-u)^\alpha F(u)$, where $F$ is regular in $u$ at $u=1$. The indicial exponents near the horizon are given by
\begin{equation} \label{eq:alphaMagnetic}
\alpha = \pm \frac{i \hat{\omega}}{2 u_1} \, ,
\end{equation}
of which we choose the solution with the negative sign as it corresponds to infalling waves. We solve the fluctuation equation for $h_{xy}$ and require the solution to vanish at the AdS boundary $u=0$, which gives us the tensor quasinormal frequencies for the two parameter family of magnetic branes (depending on the parameters $b$ and $u_1$).

For $\hat{k}=0$, linearizing the Einstein-Maxwell equations in the fluctuations, we also find six coupled equations for the four scalar fluctuations $h_{tt},\, h_{xx},\, h_{yy},\, h_{zz}$. These six equations decouple from all other metric and gauge field fluctuations, and can be found explicitly in the accompanying notebook~\cite{notebookLink}. From them a single equation of motion for  $(h_{zz}-h_{xx})$ can be derived. Due to the rotational symmetry in the $xy$-plane we may set $h_{yy}\equiv 0$ without loss of generality. It is convenient to work with one raised index on the metric perturbations, i.e. we work with $h_t^t,$ and $h_x^x,$ while $h_z^z$ is eliminated by making the field redefinition $h^z_z\equiv \chi+h^x_x$. 
Using the $tt,\, tu$, and $uu$ components of the Einstein equations $h^x_x$ can be eliminated. Making use of the background equations for U, V, and W, given by \eqref{eq:UVW}, we eliminate $U'', \, V'',\, W''$, and $W'$, which also leads to cancellation of all terms involving $h^t_t$ without any radial derivative acting on it. Lastly, the $xx$, $yy$, and $zz$ components of Einsteins can be solved for ${h^t_t}'', {h^t_t}'$, and a differential equation involving only the desired combination of fields, $\chi=h^z_z-h^x_x$, which takes the form 
\begin{eqnarray}\label{eq:scalarMagnetic}
0 & = & \chi''(u) + a(u, \hat{\omega}) \chi'(u) + b(u,\hat{\omega}) \chi(u) \, ,
\end{eqnarray}
where the coefficients $a$ and $b$ diverge at the horizon as usual, and depend on the radial coordinate $u$ as well as on the frequency $\hat\omega$. Their explicit form can be found in \cite{notebookLink}.

Finally, the scalar quasinormal modes are obtained by again demanding the Dirichlet condition $\chi(u=0) = 0$ at the $AdS$-boundary, and the infalling boundary condition at the horizon
\begin{equation}
\chi(u) = (1-u)^{-\frac{i\hat\omega}{2u_1}} H(u) \, ,
\end{equation}
where $H$ is a regular function of $u$ at $u=1$.

%%%%%%%%%%%%%%%%%%%%%%%%%%%%%%%%%%%%%%%
\section{Computational methods} \label{sec:methods}

In this section we review the two methods we use in order to find quasinormal modes (QNMs). 

%--------------------------------------------------------
\subsection{Shooting}

The shooting method has been applied in many cases for finding quasinormal modes, see for example~\cite{Kaminski:2009ce,Kaminski:2008ai}. Any fluctuation $\phi$ (e.g. a metric component, gauge field, scalar fluctuation) generically satisfies an equation of motion which is of second order in the radial coordinate, see equations \eqref{eq:hxy} and \eqref{eq:hxyMagnetic} for our specific cases at hand. Hence, one has to specify two boundary conditions in order to numerically solve these equations.\footnote{Fluctuations can couple to each other, see \cite{Amado:2009ts,Kaminski:2009dh} for a systematic method for finding quasinormal modes of such coupled systems of fluctuation equations.} The basic idea of the shooting method is to specify these two boundary conditions at the horizon, and then adjust the free parameter given by the frequency $\omega$ to find the desired solution at the boundary. One of these two horizon boundary conditions is the infalling boundary condition. This amounts to picking the negative sign in equations \eqref{eq:nearHorizonhxy} and \eqref{eq:alphaMagnetic}. The remaining boundary condition is a mere normalization coefficient. In order to ensure that the solution is a quasinormal mode one has to vary the frequency $\omega$ until the field vanishes $\phi(r_B) = 0$ at the boundary $r=r_B$. This can be efficiently achieved using a numerical optimization procedure such as Mathematica's FindRoot~\cite{Wolfram}, nesting this FindRoot with the NDSolve that finds the numerical solution, see~\cite{notebookLink} for more details. It is now clear that the shooting method depends crucially on the convergence properties of the horizon expansion and the precision of the initial values provided from evaluating that horizon expansion numerically at a numerical cut-off $r=r_{cut-off}$ near the true horizon $r_H$.

Once a QNM is found at a particular point in parameter space, e.g. specified by $\tilde q$, we can easily obtain QNMs for an $\epsilon$-neighborhood around this initial point in parameter space, i.e. $\tilde q +\epsilon$. We merely give the previous QNM frequency to FindRoot as a best estimate of the result, and find the corresponding QNM at that neighboring point in parameter space. Generally, the performance increases when $\epsilon$ is decreased. However, there is only a limited region in parameter space which is accessible to this method without increasing the working precision, or improving performance of the algorithm otherwise, as we discuss in the next paragraph.

One advantage of the shooting method is that it can be used for finding quasinormal modes of spacetimes that are only known numerically. While the method is also quite fast, it still has its limitations. One such limitation is that performance decreases rapidly as quasinormal frequencies with large imaginary part (Im$\omega \gg T$) are considered. See~\cite{Kaminski:2009ce} for a discussion and example calculations for this and related issues. In order to overcome this limitation one has to compute more coefficients of the near horizon expansion in order to be able to provide initial values further away from the horizon location $r_H$ (without leaving the radius of convergence of the horizon expansion). 

%--------------------------------------------------------
\subsection{Continued fractions}

The continued fraction method has also been put to great use in numerically determining quasinormal modes, as in~\cite{Starinets:2002br,Janiszewski:2011sx}. This method makes use of an elegant mathematical theorem originally due to Pincherle, and later generalized in~\cite{Parusnikov}. Instead of dealing directly with the differential equation of motion, i.e. \eqref{eq:hxy}, this technique works with the power series solution expanded about a singular point, such as the horizon. Generically, the radius of convergence of such an expansion is limited by the distance to the next nearest singular point of the differential equation. On the other hand, Pincherle's theorem, and its generalization, give a criterion to determine when this radius of convergence is increased. 

An increased radius of convergence allows the calculation of quasinormal modes for the following reason. The fluctuation's equation of motion \eqref{eq:hxy} has a regular singular point at the boundary, where it's characteristic exponents are $\Delta_{\pm}=\pm 2$. This means that a solution there will generically behave as
\begin{eqnarray}
h_{xy}&=&a_0 r^2+a_1 r+\cdots+a_4 r^{-2} \log r+\cdots\nonumber\\
&+&b_0 r^{-2}+b_1 r^{-3}+\cdots,
\label{eq:bdryexp}
\end{eqnarray} 
where $a_0$ and $b_0$ are the two free constants determining the expansion and, importantly, since the characteristic exponents differ by an integer, the logarithmic term starting with $a_4$ is needed so that the two solutions are linearly independent. The other $a_i$ and $b_i$ are determined in terms of $a_0$ and $b_0$ by the equation of motion, and crucially $a_4$ vanishes only when $a_0$ is zero.
Therefore, by applying Pincherle's theorem to the power series of the infalling solution at the horizon, we can determine a criterion on the frequency $\omega$ such that this series is convergent at the boundary\footnote{This method fails when the boundary is not the next nearest singular point to the horizon.}. Converging to an analytic function, it cannot have the logarithmic term $a_4$, which in turn means that this solution has $a_0=0$. Such $\omega$ therefore determine which infalling solutions have the leading behavior $h_{xy}\sim b_0 r^{-2}$ as $r\to\infty$, that is, they are Dirichlet at the boundary: they are the quasinormal frequencies. 

\subsubsection{Pincherle's theorem}\label{sec:pin}

Pincherle's theorem, and its generalization, relates a recursion relation, such as that coming from a power series solution of a differential equation about a singular point, to a continued fraction. The convergence of the continued fraction corresponds to the existence of a minimal solution of the recursion relation, where we recall that a sequence $h_n$ is minimal if
\begin{eqnarray}
\lim_{n \to \infty}\frac{h_n}{g_n}=0,\nonumber 
\label{eq:minimal}
\end{eqnarray} 
for all other sequences $g_n$ satisfying the recursion relation. A minimal solution implies an increased radius of convergence of the power series solution, beyond the next singular point. For a full proof see the reference~\cite{Parusnikov}, the important results are collected in Appendix \ref{app:Pincherle}.  

The continued fraction method of calculating the quasinormal frequencies is useful as a numerical check of other methods. Beyond that it is powerful because it is often faster than the shooting method, and can explore further into the complex plane with greater accuracy. Its major shortcoming is that it can only be applied to find the quasinormal modes of fluctuations around analytically know backgrounds, and for that reason is inapplicable to the magnetic branes also of concern at present.

% ............................
\subsubsection{Application to Reissner-Nordstr\"om}\label{sec:ex} 

In applying the continued fraction method to the calculation of quasinormal modes, the first thing needed is an expansion of the equation of motion about a singular point. The desired singular point is the horizon, as that allows the infalling nature of the perturbation to be made directly in the ansatz. We will start with the background metric of the Reissner-Nordstr\"om black brane in the coordinates of \eqref{eq:zRNMetric}. The perturbation ansatz $g_{xy}=e^{(-i \omega t+ i k z)}u h_{xy}(u)$ is made and the Einstein-Maxwell equations are linearized in $h_{xy}$, as in Section \ref{sec:RNfluc}. It is useful to work with the rescaled quantities $\tilde{\omega}\equiv\omega/r_H$ and $\tilde{k}\equiv k/r_H$ as this eliminates $r_H$ from the equation of motion. One last change of radial coordinate is made: $u\equiv 1-\tilde{r}(2-\tilde{q}^2)^2$. This is done as the horizon is now at $\tilde{r}_H=0$, which allows a cleaner series expansion there. The charge dependent rescaling simply helps with numerical stability in the region of extremality, $\tilde{q}\to\sqrt{2}$.

The equation of motion for $h_{xy}$ has regular singular points at the horizon and the boundary $\tilde{r}_b=1/(2-\tilde{q}^2)^2$, as well as at the radial coordinates\footnote{The location $\tilde{r}_+$ is always further from the horizon than the boundary is, but at $\tilde{q}\geq \sqrt{3}/2$ the singular point $\tilde{r}_-$ is the next nearest singular point to the horizon. 
This corresponds to charges greater than $91.7\% q_{ex}$ for the unscaled charge $q$, and invalidates the continued fraction method in this region, as mentioned above.} $\tilde{r}_\pm\equiv (-1+2\tilde{q}^2\pm\sqrt{1+4\tilde{q}^2})/(2\tilde{q}^2(2-\tilde{q}^2)^2)$. Determining the leading behavior near the horizon, the characteristic exponents are $\alpha=\pm \imath \tilde{\omega}/(2(2-\tilde{q}^2))$, with the lower sign corresponding to the desired infalling condition. Peeling off this factor, i.e. $h_{xy}(\tilde{r})\equiv\tilde{r}^{\alpha}f(\tilde{r})$, the equation of motion can be written as
\begin{equation}
0=f(\tilde{r})\sum\limits_{i=0}^4 r_i \tilde{r}^i+f'(\tilde{r})\sum\limits_{i=0}^5 s_i \tilde{r}^i+f''(\tilde{r})\sum\limits_{i=1}^6 t_i \tilde{r}^i,
\label{eq:eomf}
\end{equation} 
where $r_i$, $s_i$, and $t_i$ are coeffcients that depend on $\tilde{q}$, $\tilde{\omega}$, and $\tilde{k}$ and can be found in the notebook~\cite{notebookLink}.

Postulating a series solution, $f(\tilde{r})\equiv\sum\limits_{i=0}^\infty c_i \tilde{r}^i$, the coefficients $c_i$ must obey the recursion relation
\begin{equation}
c_n=-\sum\limits_{i=1}^5 c_{n-i}\left(\frac{r_{i-1}+(n-i)s_i+(n-i)(n-i-1)t_{i+1}}{ns_0+n(n-1)t_1}\right),
\label{eq:rr}
\end{equation}
with the initial conditions
\begin{align}
c_0 &= 1,\quad c_1=-\frac{c_0 r_0}{s_0},\quad c_2=-\frac{c_1(r_0+s_1)+c_0r_1}{2s_0+2t_1},\nonumber\\ 
c_3 &= -\frac{c_2(r_0+2s_1+2t_2)+c_1(r_1+s_2)+c_0r_2}{3s_0+6t_1},\nonumber\\ 
c_4 &= -\frac{c_3(r_0+3s_1+6t_2)+c_2(r_1+2s_2+2t_3)+c_1(r_2+s_3)+c_0r_3}{4s_0+12t_1}.
\label{eq:cinit}
\end{align}
Comparing to Parusnikov's Theorem 2, see Appendix \ref{app:Pincherle}, this 6 term recursion relation implies $m=4$, and the 5-dimensional vector $\vec{p}_n$ needed to construct the continued fraction has components
\begin{equation}
p_{i,n}=-\frac{r_{5-i}+(n+i-6)s_{6-i}+(n+i-6)(n+i-7)t_{7-i}}{ns_0+n(n-1)t_1}.
\label{eq:vecp}
\end{equation}

The desired continued fraction $\vec{f}_0$ is now defined by the limit given in \eqref{eq:f0}. In practice, one calculates $\vec{f}_0^{\,(i)}$ for large $i$; changing this parameter allows one to understand the convergence properties of the limit, and obtain any desired accuracy. The final step in calculating the quasinormal frequencies is to demand that a minimal solution to the recursion relation \eqref{eq:rr} exists by finding the values of $\tilde\omega$ such that Theorem 7/8 of Appendix \ref{app:Pincherle} is obeyed for the initial conditions \eqref{eq:cinit}. Explicitly, we find the solutions of
\begin{equation}
c_4=\left(\frac{-1}{f_{4,5}^{(i)}}\right)c_0+\sum\limits_{j=1}^3\left(\frac{-f_{j,5}^{(i)}}{f_{4,5}^{(i)}}\right)c_j,
\label{eq:qnfeq}
\end{equation}  
which is a high order polynomial in $\tilde\omega$. Note that  $\vec{f}_5$ is defined in the appendix (below Theorem 7/8). Figure \ref{fig:conv} shows the movement of the first three quasinormal modes calculated in this way as the number of iterations $i$ is increased from 3 to 41, in steps of two. The modes generically move up and to the right, and converge rather quickly. In practice, 50 iterations were used to obtain the $10^{-6}$ accuracy in the continued fraction data quoted throughout. 
\begin{figure}[]
\includegraphics[height=7cm]{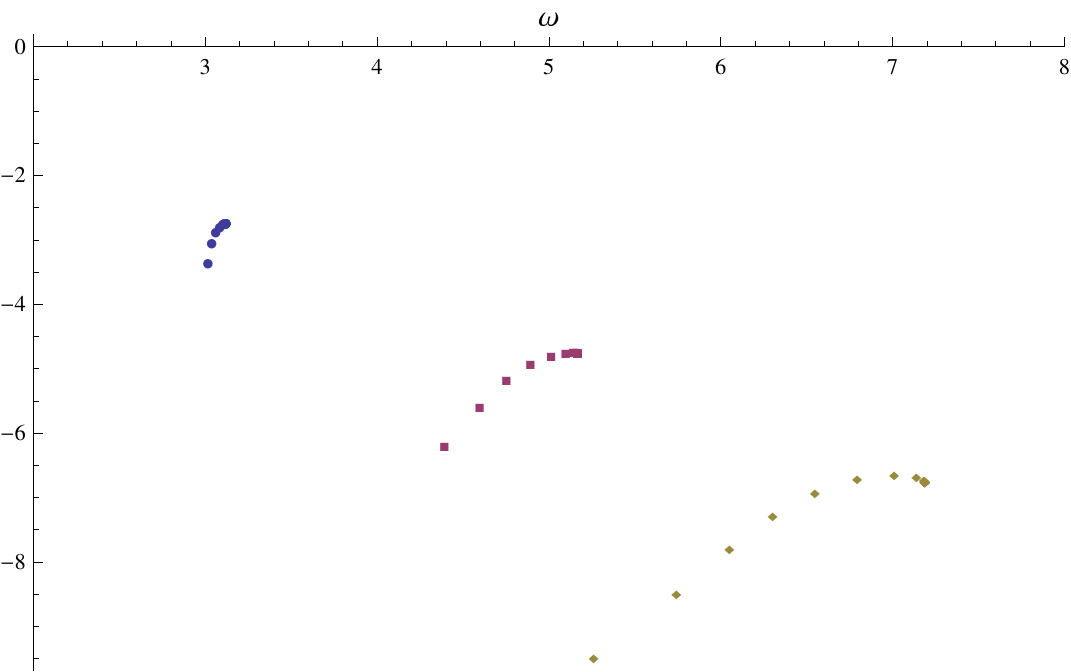}
\caption{\label{fig:conv}
{\it Convergence of the continued fraction method.} The first three quasinormal modes (circles, squares, diamonds) at $\tilde{q}=\tilde{k}=0$ are plotted in the complex frequency plane as the number of iterations is increased. 
}
\end{figure}

%%%%%%%%%%%%%%%%%%%%%%%%%%%%%%%%%%%%%%%
\section{Results} \label{sec:results}

In this section we describe the quasinormal modes (QNMs) which are present in the tensor and vector fluctuations around Reissner-Nordstr\"om black branes, and in the tensor fluctuations around magnetic black branes. In the latter case we also study the scalar quasinormal modes. We discuss the behavior of the quasinormal modes as a function of the relevant parameters and provide explicit data for various points in parameter space for comparison. Throughout the present section we employ different units: we report quasinormal frequencies in units of mass density, by which we mean that we display values in tables and figures which are computed at the parameter value $m=1$, if not specified otherwise. Where indicated, we display the frequencies in units of temperature $(\pi T)$, or of mass density after a rescaling with the inverse horizon radius, $\omega \to \omega/r_H$ (stemming from a rescaling of the time coordinate $t \to r_H t$ in the metric~\eqref{eq:zRNMetric}).\footnote{Note that in the coordinates discussed in this sentence the energy density is now temperature and charge dependent.} The latter values are referred to as ``rescaled QNMs". The charge density is given either in units of mass density, i.e. $q$ in units of $(m/3)^{3/4}$, or in dimensionless values denoted by $\tilde q$. Momenta $k$ are given in units of mass density $m$, momenta $\tilde k$ and $\hat k$ are the rescaled versions.
 
%--------------------------------------------------------
\subsection{Quasinormal frequencies of Reissner-Nordstr\"om black branes}
We find two distinct sets of modes, see the schematic plot in figure \ref{fig:QNMSets}. The first set (set 1) also exists at zero charge density and has been analyzed in~\cite{Starinets:2002br}. The second set of modes (set 2) consists of purely imaginary quasinormal modes. We observe the latter to approach the origin at and above values of $\tilde q = 0. 5 \tilde q_{ex}$. These purely imaginary modes of set 2 are not visible at all at small or zero charge densities, as they are lying far down the imaginary axis.

In agreement with previous results~\cite{Starinets:2002br}, we find a tower of quasinormal modes with frequencies that have nonzero real and imaginary part. The values obtained from our continued fraction method are in very good agreement with those obtained from the shooting method (8 significant digits agree), and both are in good agreement with previous results found by Starinets. These results are computed at vanishing charge density. This agreement is a non-trivial result, as our continued fraction calculation is quite different from that of Starinets~\cite{Starinets:2002br}: ours is a 4 dimensional vectorial continued fraction, while~\cite{Starinets:2002br} uses a (1 dimensional) scalar continued fraction. See table \ref{tab:QNMsRN_consistency}. 
\begin{table}[]
\begin{tabular}{|c|c|c|c|}
\hline
mode \#  & shooting & cont. fraction and Starinets  \\ \hline
1  &  $\pm 3.1194516 - i 2.7466757$ &  $\pm 3.119452 - i 2.746676 $ \\
2  &  $\pm 5.1695210 - i 4.7635701$ &  $\pm 5.169521 - i 4.763570$ \\
3  &  $\pm 7.1879308 - i 6.7695650$ &  $\pm 7.187931 - i 6.769565$ \\
4  &  $\pm 9.1971992 - i 8.7724814$ &  $\pm 9.197199 - i 8.772481$  \\
5  &  $\pm 11.2026887 - i 10.7740258$ &  $\pm 11.202676 - i 10.774162 $ \\ \hline
\end{tabular}
\caption{ \label{tab:QNMsRN_consistency}
{\it Consistency with previous data.} The values of the five lowest lying quasinormal mode frequencies $\tilde \omega$ for the tensor fluctuations, e.g. $h_{xy}$, are shown in units of mass density. The first column labels the mode number $n=1,\, 2,\, 3,\, 4,\, 5$. The second column shows the quasinormal mode frequencies obtained from the shooting method, the third column shows those obtained from our continued fraction method which is in complete agreement with the values previously obtained by Starinets in \cite{Starinets:2002br}. These results are computed at vanishing charge density.}
\end{table}

\begin{figure}[]
\includegraphics[height=8cm]{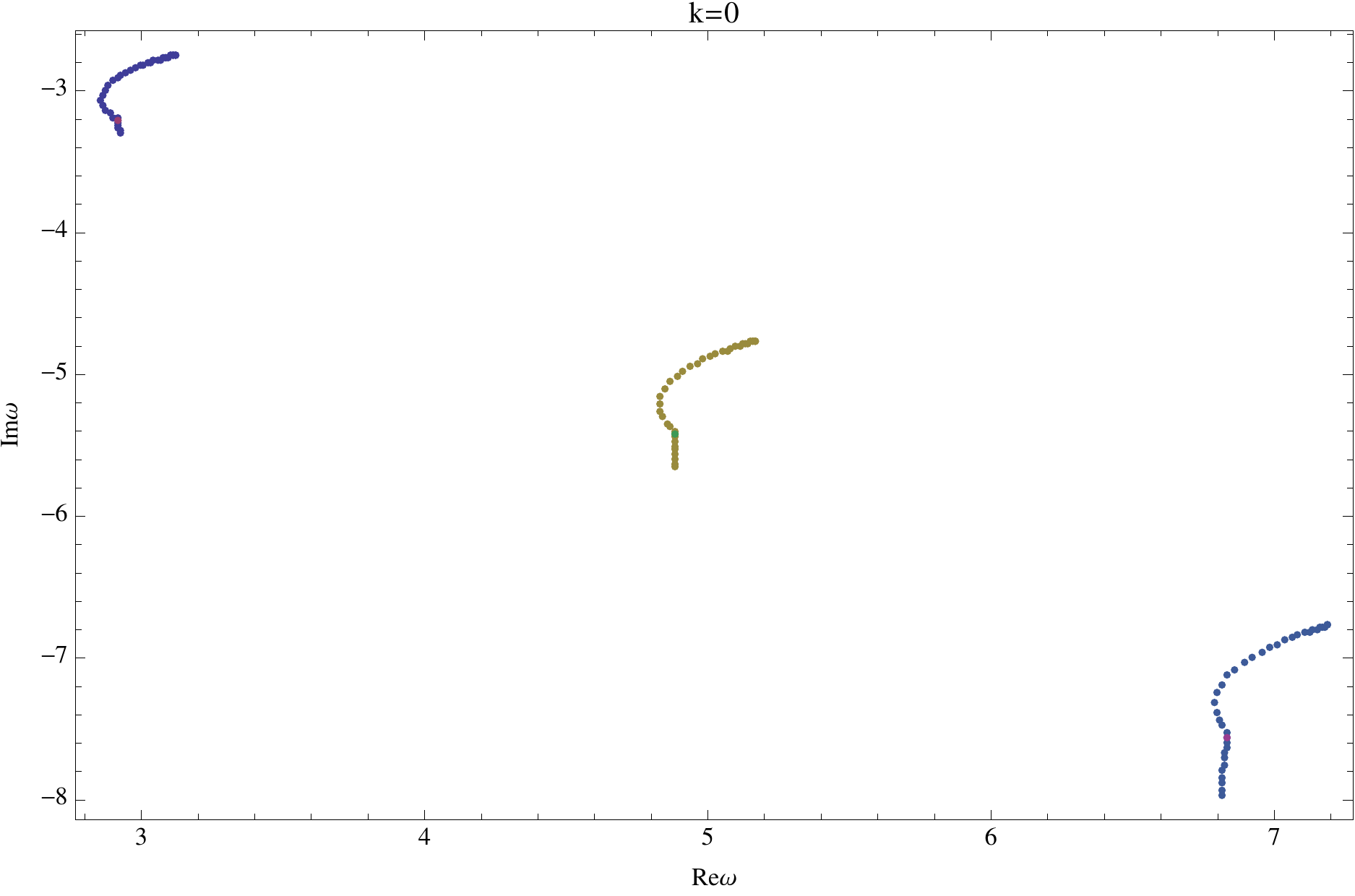}
\caption{\label{fig:QNMsRN_changeq_k0}
{\it Reissner-Nordstr\"om QNMs of $h_{xy}$ at $q\neq 0$, $k=0$.}
The three trajectories of the three lowest quasinormal modes are shown as the momentum is fixed to $k=0$ and the charge is increased from $q=0$ to $q= 0.9 \, q_{ex}$ in increments of $\Delta q=q_{ex}/40$. The uppermost point in each trajectory (top right point) corresponds to vanishing charge $q=0$. For increasing charge, the modes each move to more negative imaginary frequency values.
}
\end{figure}

At increasing charge density and fixed momentum $k=0$, the modes of set 1 move as indicated in figure~\ref{fig:QNMsRN_changeq_k0}, displayed in the frequency $\omega$. There is an obvious kink in the trajectories at larger charge densities $\tilde q\approx 0.6 \tilde q_{ex}$. 
However, plotting the QNMs in the rescaled frequency variable $\tilde\omega$, no kinks are visible. Only upon multiplying $\tilde \omega$ with the horizon radius, those kinks appear in the QNM trajectories. These kinks appear suspiciously close to the region where convergence of the horizon expansion within our shooting method gets increasingly worse.\footnote{This is caused by fact that the inner horizon, located at $r_-$, is approaching the outer horzion, located at $r_H$. 
The kink in the quasinormal mode trajectories appears where the difference $|r_H-r_-|$ is comparable to the horizon cut-off located at $r_{cut-off}$, i.e. $|r_H-r_- |\approx |r_H-r_{cut-off}|$.} However, we have checked that the location of the kinks does not depend on the numerical horizon and boundary cut-offs that we choose. Independently, the continued fraction method confirms the kinks in the trajectories at the same locations. That said, when plotting the frequencies in units of temperature ($\pi T$), or in units of $r_H$, we observe no kinks whatsoever, merely smooth trajectories. Hence, these kinks should be regarded as an artifact of expressing frequencies in units of mass density. This point is explored in more depth in one of the accompanying notebooks~\cite{notebookLink}.

At vanishing momentum, $k=0$, the leading QNM of the metric shear fluctuation $h_{xy}$ is listed for increasing charge $\tilde q$ of the Reissner-Nordstr\"om black brane in table~\ref{tab:QNMsRN_k0}. The right column shows the frequency values in units of $(\pi T)$, while the left column shows the rescaled values which we obtain directly from our numerical procedure because therein we have worked with conveniently rescaled quantities, as seen from the metric~\eqref{eq:zRNMetric}. Those rescaled coordinates are convenient for our near-equilibrium setup. For a comparison to the late-time behavior of physical quantities obtained from numerical non-equilibrium codes the values in the right column are more suitable. In comparison with~\cite{Fuini:2015hba}, our lowest QNM values agree up to deviations of 0.2 percent, as can be seen from Table 1 in~\cite{Fuini:2015hba}.
\begin{table}[]
\begin{tabular}{|c|c|c|}
\hline
$q/q_{ex}$ & $\tilde \omega$ (rescaled) & $\omega/(\pi T)$ \\
\hline 
0    & {$\pm 3.11945-  i 2.74668$} & {$\pm3.11945 - i 2.74668$} \\
0.1 & {$\pm3.11652  -i 2.75222$} & {$\pm3.12256 - i 2.75756$} \\
0.2 & {$\pm3.10759  -i 2.76953$} & {$\pm3.13227 - i 2.79153$} \\
0.3 & {$\pm3.09236  -i 2.80090$} & {$\pm3.14994 - i 2.85305$} \\
0.4 & {$\pm3.07057 - i 2.85111$} & {$\pm3.17859 - i 2.95141$} \\
0.5 & {$\pm3.04312 - i 2.92958$} & {$\pm3.22533 - i 3.10499$} \\
0.6 & {$\pm3.01749 - i 3.05496$} & {$\pm3.31032 - i 3.35142$} \\
0.7 & {$\pm3.03304 - i 3.25409$} & {$\pm3.50619 - i 3.76173$} \\
0.8 & {$\pm3.16481 - i 3.47084$} & {$\pm3.9918 - i 4.3778$} \\
0.9 & {$\pm3.32234 - i 3.68357$} & {$\pm5.0231 - i 5.56924$}\\
\hline
\end{tabular}
\caption{ \label{tab:QNMsRN_k0}
{\it Reissner-Nordstr\"om QNMs of $h_{xy}$ at $q\neq 0$, $k=0$.}
Lowest QNM frequency in set 1 for increasing charge densities $q$ (in units of $(m/3)^{3/4}$) given in fractions of the extremal charge density $q_{ex}$. The left column provides the QNM frequencies rescaled with a factor $r_H$, while the right column shows the frequencies of the same modes in units of $(\pi T)$.
}
\end{table}

\begin{figure}[]
\includegraphics[height=7cm]{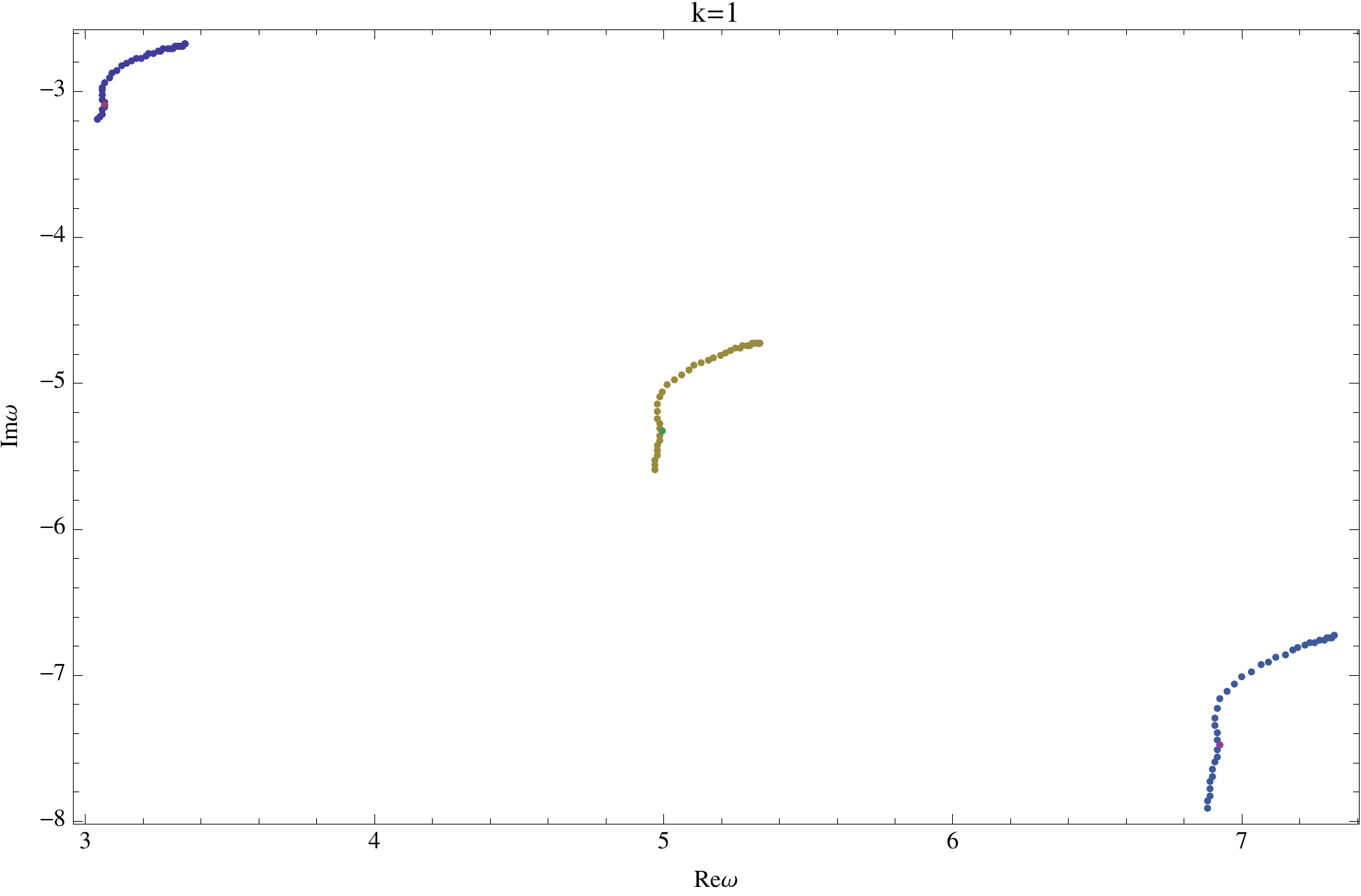}
\caption{\label{fig:QNMsRN_changeq_k1}
{\it Reissner-Nordstr\"om QNMs of $h_{xy}$ at $q\neq 0$, $k=1$.}
The trajectories of the three lowest quasinormal modes are shown as the momentum is fixed to $k=1$ and the charge is increased from $q=0$ to $q= 0.9 \, q_{ex}$ in increments of $\Delta q=q_{ex}/40$.
}
\end{figure}

\begin{figure}[]
\includegraphics[height=7cm]{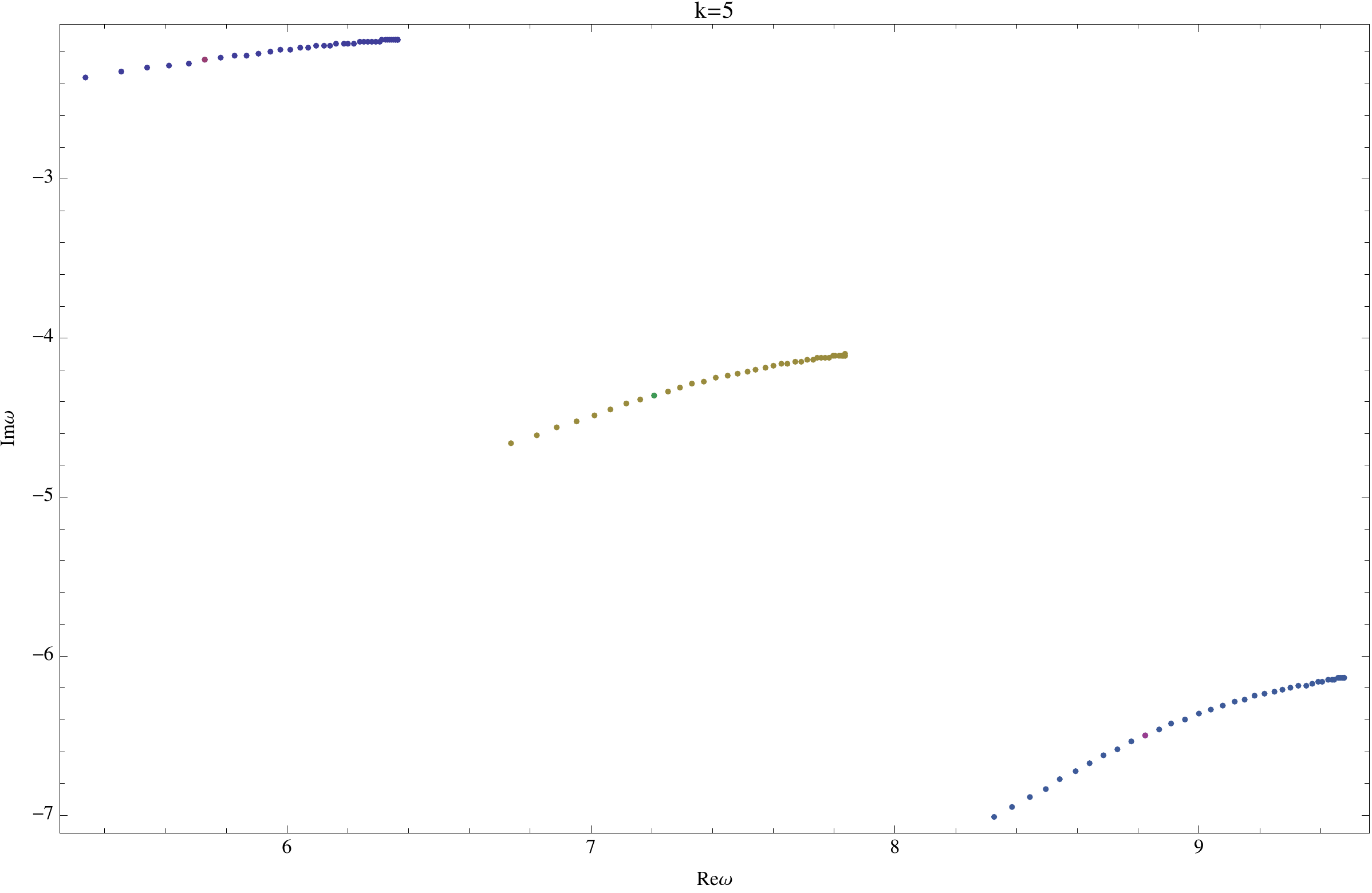}
\caption{\label{fig:QNMsRN_changeq_k5}
{\it Reissner-Nordstr\"om QNMs of $h_{xy}$ at $q\neq 0$, $k=5$.}
The trajectories of the three lowest quasinormal modes are shown as the momentum is fixed to $k=5$ and the charge is increased from $q=0$ to $q= 0.9 \, q_{ex}$ in increments of $\Delta q=q_{ex}/40$.
}
\end{figure}

\begin{figure}[]
\includegraphics[height=7cm]{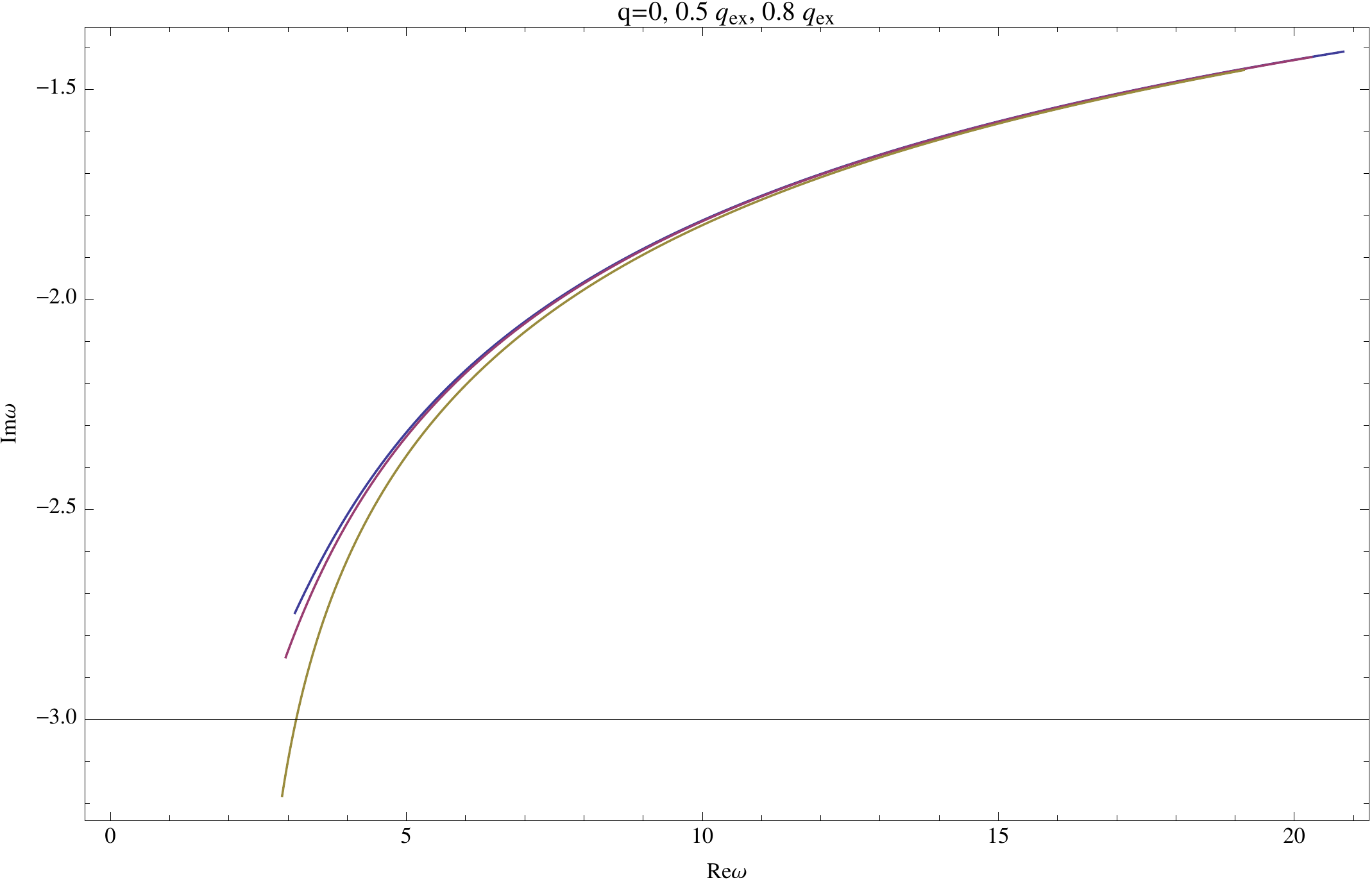}
\caption{\label{fig:QNMsRN_changek}
The trajectory of the three lowest quasinormal modes are shown as the charge is fixed to $q=0, 0.5 \, q_{ex},\,$ or $0.8 \, q_{ex} $ while the momentum is increased from $k=0$ to $k=20$ in increments of $\Delta k=0.1$.
}
\end{figure}

Fixing the momentum at larger values $k=1/2$ and $k=1$ does not change the qualitative behavior of the modes, as seen from tables~\ref{tab:QNMsRN_khalf} and~\ref{tab:QNMsRN_kone}, respectively. Also in the quasinormal mode trajectories computed at fixed nonzero momentum $k=1/2$ we observe kinks, as before at $k=0$. See figure~\ref{fig:QNMsRN_changeq_k1}. However, at large fixed momentum, $k=5$, we observe that the kinks disappear from the regime accessible to our numerical methods, as seen in figure~\ref{fig:QNMsRN_changeq_k5}. Another interesting exercise is to fix the charge density and then follow the quasinormal modes while increasing the momentum. The result of this exercise can be seen in figure~\ref{fig:QNMsRN_changek} for fixed charges $q=0, \, 0.5 q_{ex}, \,0.8 q_{ex}$ (blue, red, brown curves, respectively). At large momentum, the trajectories all asymptote to the same ``attractor'' curve, regardless of the charge. This seems reasonable because at large momentum the scale set by the charge is negligible compared to the scale set by the momentum, in other words all trajectories at large momentum should asymptote to the zero charge trajectory, which is confirmed by our data.
\begin{table}[]
\begin{tabular}{|c|c|c|c|}
\hline
mode \#  & 1 & 2 & 3  \\ \hline
$ q=0$  & $\pm3.1779741 - i 2.7290839$ & $\pm 5.2112434 - i 4.7519066$ & $\pm 7.2214147 - i 6.7604134$   \\
$q=0.25\, q_{ex}$  & $\pm3.1413918 - i 2.7474627$ & $\pm 5.1413864 - i 4.7936400$ &  $\pm 7.1185609 - i 6.8281811$ \\
$q=0.5\, q_{ex}$  & $\pm3.0262710 - i 2.8281461$ & $\pm 4.9367320 - i 4.9919741$ & $\pm 6.8493201 - i 7.1646578$  \\
$q=0.75\, q_{ex}$ & $\pm 2.9252877 - i 3.0924526$ & $\pm 4.9163769 - i 5.3955189$ & $\pm6.8500839 - i 7.6457490$ \\
\hline
\end{tabular}
\caption{ \label{tab:QNMsRN_khalf}
{\it Nonzero momentum.} Reissner-Nordstr\"om black brane quasinormal mode frequencies $\omega$ for the tensor fluctuations, e.g. $h_{xy}$, at fixed momentum $k=1/2$ and increasing charge density $q$. 
}
\end{table}
\begin{table}[]
\begin{tabular}{|c|c|c|c|c|}
\hline
mode \#  & 1 & 2 & 3 \\ \hline
$ q=0$ & $\pm 3.3440382 - i 2.6808260$ & $\pm5.3317107 - i 4.7178383$ & $\pm7.3188762 - i 6.7331804$ \\
$q=0.25\, q_{ex}$ & $\pm3.3102870 - i 2.6965339$ & $\pm5.2651313 - i 4.7562293$  &$\pm 7.2195106 - i 6.7971758$ \\
$q=0.5\, q_{ex}$ & $\pm 3.2045995 - i 2.7642384$ &$\pm 5.0692976 - i 4.9339576$ & $\pm6.9544811 - i7.1066419$ \\
$q=0.75\, q_{ex}$ & $\pm 3.0864530 - i 2.9801352$ & $\pm 5.0079788 - i 5.3207002$  &$ \pm 6.9204927 - i 7.5887749$ \\
\hline
\end{tabular}
\caption{ \label{tab:QNMsRN_kone}
{\it Nonzero momentum.} Reissner-Nordstr\"om black brane quasinormal mode frequencies $\omega$ for the tensor fluctuations, e.g. $h_{xy}$, at fixed momentum $k=1$ and increasing charge density $q$. 
}
\end{table}

\begin{figure}[]
\includegraphics[height=7cm]{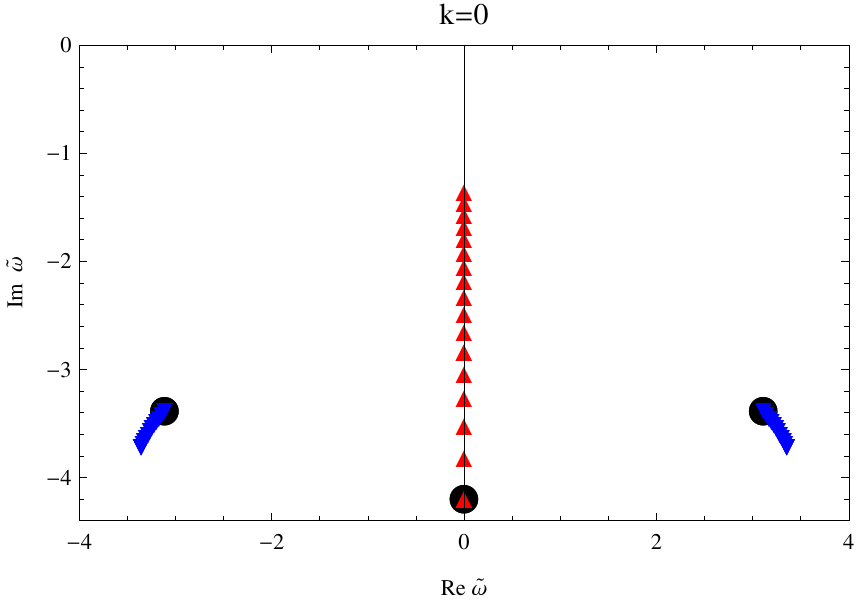}
\caption{\label{fig:QNMsRN_enterImaginarySet}
{\it At large charge densities, a purely imaginary QNM enters the low frequency regime.} We display the rescaled frequencies after $\omega\to \omega /r_H$. The trajectories of the three quasinormal modes closest to the origin $\tilde \omega=0$ are shown as the charge increases from $\tilde q=0.6$ to $1.0$ in increments of $\Delta \tilde q=1/40$. The momentum is fixed to $k=0$. The black dots mark the points of the respective trajectories with $\tilde q= 0.6$. The lowest QNM in set 1 and its complex conjugate mode move downward as the charge increases (blue downward-pointing triangles). The lowest QNM in set 2 moves upward as the charge increases (red upward-pointing triangles).
}
\end{figure}

\begin{figure}[]
\includegraphics[height=7cm]{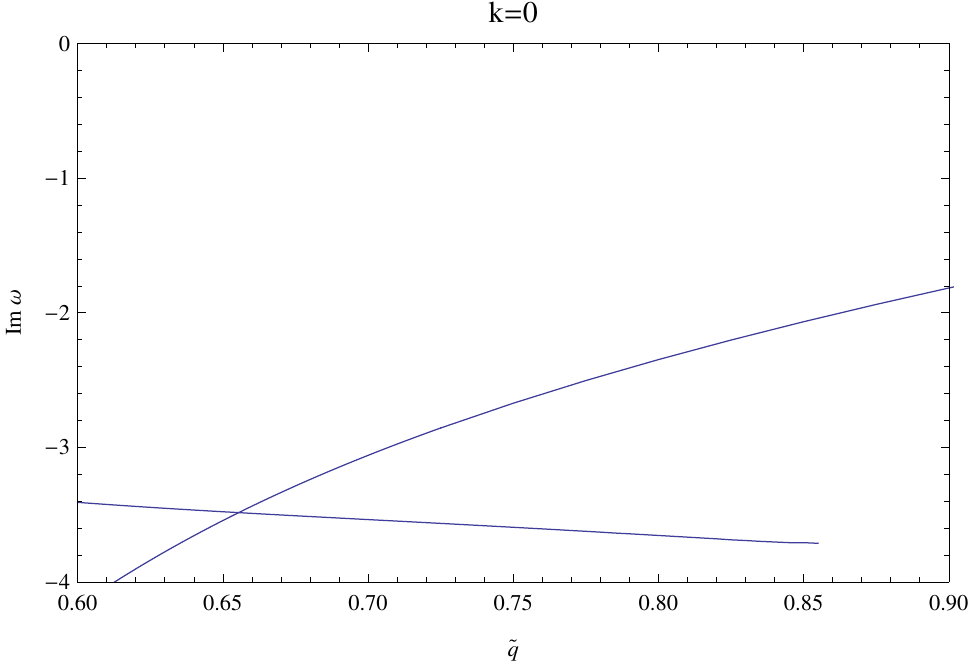}
\caption{\label{fig:QNMcrossing}
The imaginary part of the lowest QNM in set 1 is shown (line with negative slope) as a function of charge $\tilde q$. This is compared to the imaginary part of the purely imaginary QNM which is the lowest member of set 2 (line with positive slope). The two lines intersect at the charge $\tilde q_{crossing}\approx 0.655$ with an imaginary part of $\mathrm{Im}\, \omega\approx -3.50$.
}
\end{figure}

Now we turn to the second set of quasinormal modes (set 2), namely the purely imaginary ones mentioned before.\footnote{We are grateful to Michal Heller for pointing us to these kinds of modes. Further, we are grateful to Julian Sonner for discussions on the interpretation of these modes.} The member of this set 2, which is closest to the origin $\omega=0$, is shown along with the lowest QNM from set 1 (and its complex conjugate) in figure~\ref{fig:QNMsRN_enterImaginarySet}. That figure shows the trajectories of those three poles as the charge is increased from $\tilde q=0.6$ to $1.0$ in steps of $1/40$. Note that we have chosen to display the rescaled frequencies here (stemming from a rescaling of the time coordinate in the metric~\eqref{eq:zRNMetric}, as discussed before).\footnote{We also find it convenient to express the charge density in its dimensionless form here, i.e. we are using $\tilde q$ as opposed to $q$.} While the lowest mode known from set 1 moves (slowly) away from the real axis, the lowest mode of set 2 move on the negative imaginary axis towards the real axis. At a value of $\tilde q_{crossing}\approx 0.655$ the purely imaginary mode and the lowest mode of set 1 have identical imaginary parts. See figure~\ref{fig:QNMcrossing}. At charge densities $\tilde q>\tilde q_{crossing}\approx 0.655$, the lowest purely imaginary mode is the lowest of all QNMs.

As the charge density approaches its extremal value from below, the modes of set 2 are pushed towards the origin and closer towards each other. We conjecture that this may be understood as the onset of a branch cut forming along the negative imaginary frequency axis, in analogy to the near-extremal $AdS_4$ case~\cite{Edalati:2010hk,Edalati:2010pn}.\footnote{Just like in that case, at extremality, i.e. in the zero temperature limit, conformal symmetry is still broken by the nonzero chemical potential.}
Our calculations indicate that the modes of set 1 do not approach the imaginary axis in the extremal limit. 
All of the QNMs of set 1 and set 2 remain in the lower half of the complex frequency plane. In other words, the fluctuation $h_{xy}$ does not seem to cause any instabilities in the extremal limit (this statement is limited by numerical accuracy).

We have confirmed selected QNMs of set 2 with the continued fraction method to an accuracy of 4 digits.\footnote{High accuracy at all parameter values is difficult to achieve in the purely imaginary QNMs. The problem is that, at particular values of the charge $\tilde q$, the zeros of $\phi(u=0, \tilde\omega; \tilde q)$ move through poles of the function $\phi(u=0, \tilde\omega; \tilde q)$. Those poles are located at $\tilde \omega=i n (\tilde q^2-2)$ (with integer $n$), where the continued fraction method is known to yield false QNMs~\cite{Starinets:2002br,Janiszewski:2011sx}.} Various checks of both of our numerical methods indicate that these purely imaginary modes of set 2 are a robust feature which is not an artifact of the numerical methods. Consequently, the QNMs of set 2 are dominating the late-time physics of the dual field theory at fairly sizable temperatures, namely at charges $\tilde q>\tilde q_{crossing}\approx 0.655$, corresponding to temperatures $T < T_{crossing} \approx 0.228$.


Our lowest tensor QNM frequency from set 1 agrees very well with the QNMs extracted from the late-time behavior of the fully time dependent setup at the corresponding charge densities. Table 1 in~\cite{Fuini:2015hba} shows a maximum deviation of 0.2\% at fairly large charge densities. At charge densities much smaller than the extremal one, we find much better agreement. Note that Table 1 in~\cite{Fuini:2015hba} only probes charge densities for which the lowest QNM of set 1 still dominates the late-time behavior. The authors of~\cite{Fuini:2015hba} use the charge density $q$ which takes its extremal value at $\sqrt{2} (m/3)^{3/4} \approx 0.6204$ with the mass parameter $m=1$, while the extremal charge value in the tilded coordinates is $\tilde q=\sqrt{2}$. The relation between the two charges is $q=\tilde q (m/(1+{\tilde q}^2))^{3/4}$. Therefore, 90\% of extremal charge in~\cite{Fuini:2015hba} is identical to 58\% of extremal charge in the tilded coordinates. The authors of~\cite{Fuini:2015hba} report numerical difficulties with larger charge density values, whereas it is straightforward for us to reach more than 70\% of extremal charge in the tilded coordinates.

\begin{figure}[]
\includegraphics[height=7cm]{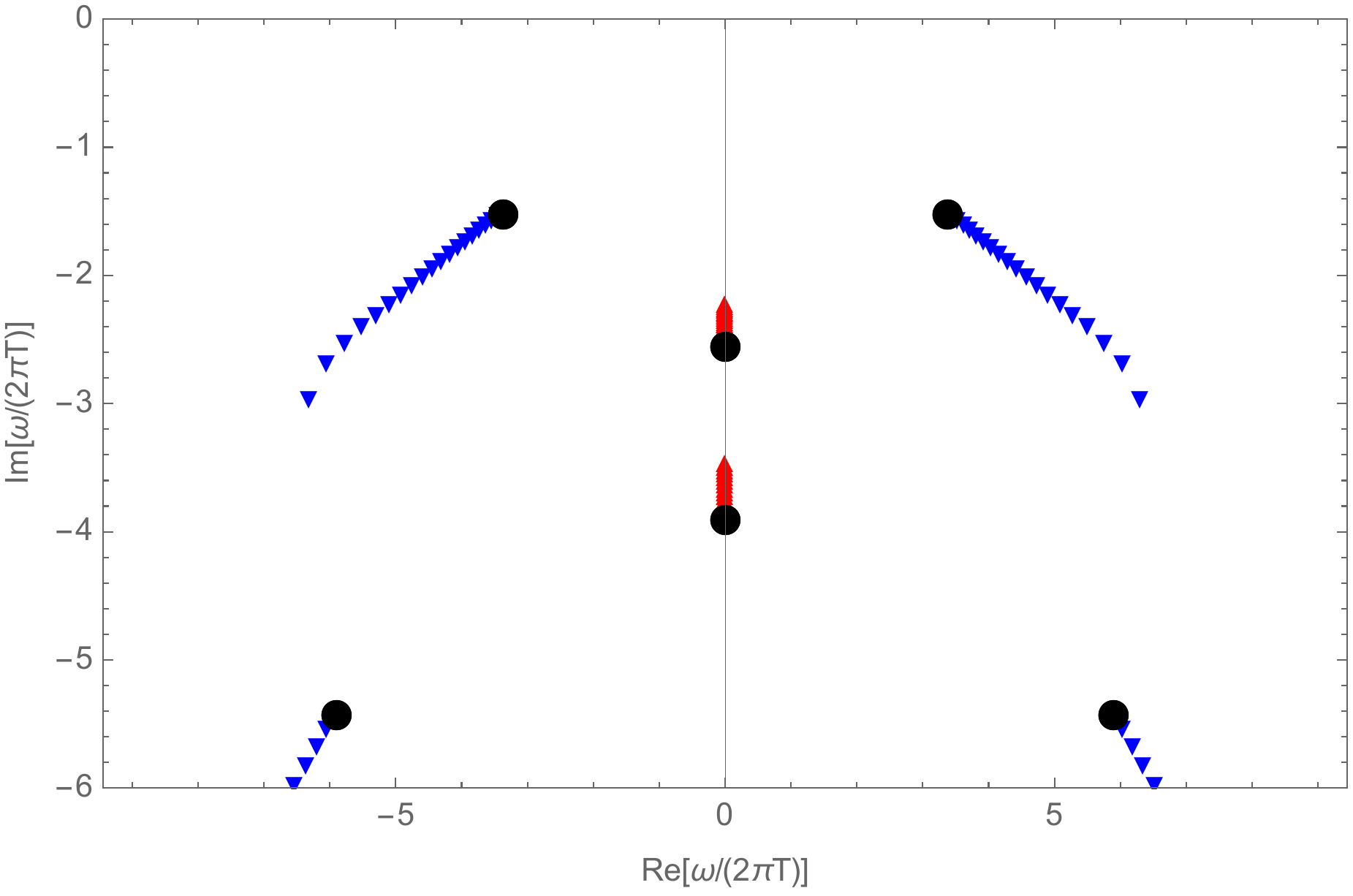}
\caption{\label{fig:vectorQNMs} 
The two lowest complex and two lowest purely imaginary RN $AdS_5$ vector QNM frequencies in units of temperature, i.e. $\omega/(2\pi T)$, are shown for increasing charge densities $\tilde q/\tilde q_{ex} = 0.800, 0.805, 0.810, \dots,\, 0.895,\, 0.900$. Like the tensor QNMs, also these vector QNMs fall into two sets. Set 1 (blue down triangles) moves down in the complex plane as the charge density is increased. Set 2 (red up triangles) moves up along the imaginary axis as charge density is increased. The black dots mark the QNMs at $\tilde q/\tilde q_{ex} = 0.800$.
}
\end{figure}

For comparison to~\cite{Maeda:2006by}, here we also compute the vector modes. The two lowest vector QNMs (members of Set 1) at vanishing momentum $\vec{k}=0$ are given for various charges in table~\ref{tab:vectorQNMs}. Figure~\ref{fig:vectorQNMs} shows their trajectories with increasing charge density, and we have chosen the same units for our frequency as the authors of~\cite{Maeda:2006by}, i.e. we have plotted $\omega/(2 \pi T)$. Note that the authors of~\cite{Maeda:2006by} display results at nonzero momentum only, while all of our vector QNM results are at vanishing momentum. Comparing to figure~1 from~\cite{Maeda:2006by}, we observe qualitative agreement to the extent allowed by the visual comparison~\footnote{The figure in~\cite{Maeda:2006by} shows vector QNMs at nonzero spatial momentum. An analysis of the vector modes at nonzero momentum is beyond the scope of this work, hence we content ourselves with a qualitative comparison.}. We compare to the left plot in figure~1 of~\cite{Maeda:2006by}. In light of our own results, we interpret their green data set ($r_-/r_+=0.5$) as similar to our vector QNMs at intermediate charge values, i.e. roughly half of the extremal value of $\tilde q$. In that case, there is already one mode of Set 2 visible, sitting on the imaginary axis in their figure~1. We interpret their blue data set ($r_-/r_+=0.95$) as showing only modes of Set 2 whereas the modes of Set 1 are outside the plot range of their figure~1. Our numerical data at vanishing momentum does not show the hydrodynamic (diffusion) mode closest to $\omega/(2\pi T)=0$ indicated by the box in figure~1 of~\cite{Maeda:2006by}.
 At vanishing charge $\tilde q=0$, we recover the values, $\omega/(2\pi T) = n (\pm1-i)$ with $n=0,\, 1\, , \dots$, already obtained analytically in~\cite{Nunez:2003eq}. Just like for the tensor fluctuation, also in this vector fluctuation a second set of purely imaginary QNM (Set 2) wanders up the imaginary axis. And like the purely imaginary tensor QNMs, at a critical charge value also these imaginary vector QNMs become more relevant to the late time behavior than the modes of Set 1. In summary, the vector channel QNMs show exactly the same structure and qualitative behavior as the tensor QNMs.
\begin{table}[]
\begin{tabular}{|c|c|c|c|}
\hline
mode \#  & ${{\tilde q}/{\tilde q_{ex}}=0}$& ${{\tilde q}/{\tilde q_{ex}}= 0.8}$  & ${{\tilde q}/{\tilde q_{ex}}= 0.9 }$  \\ \hline 
\multicolumn{4}{|c|}{Set 1 (complex QNMs)} \\ \hline
1  & $\pm 1.0000000- i 1.0000000  $  & $\pm 3.3687082 - i 1.5042378$ & $\pm 6.3036022 - i 2.9733044$   \\
2  & $\pm 2.0000000- i 2.0000000$  & $\pm 5.8977388 - i 5.4108751$ & $\pm 9.0680850 - i 6.1472415$   \\ \hline
\multicolumn{4}{|c|}{Set 2 (purely imaginary QNMs)} \\ \hline
1*  & not accessible  & $\pm 0  - i 2.5337209$ & $\pm 0 - i 2.3875204 $   \\
2*  & not accessible  & $\pm 0  - i 3.8897030$ & $\pm 0 - i 3.4700708 $   \\ \hline
\end{tabular}
\caption{ \label{tab:vectorQNMs}
{\it RN $AdS_5$ vector QNMs.} The values of the two lowest lying complex and two lowest lying purely imaginary quasinormal mode frequencies in units of temperature, i.e. $\omega/(2\pi T)$, for the vector fluctuations are shown at selected charge densities $\tilde q$.}
\end{table}
%

%--------------------------------------------------------
\subsection{Quasinormal frequencies of magnetic black branes}
Here we report quasinormal mode frequencies of the magnetic black branes discussed in Section~\ref{sec:magBranes}. On the field theory side this setup allows us to examine the late-time behavior of neutral strongly coupled plasma in an external magnetic field being sheared and then relaxing. The shear occurs between the $xy$-plane and the $z$-direction. On the gravity side we have introduced scalar fluctuations $h_{tt},\, h_{xx},\, h_{yy},\, h_{zz}$ at vanishing momentum, which have common QNMs determined by the fluctuation equation \eqref{eq:scalarMagnetic}. We also study the QNMs of the tensor fluctuation $h_{xy}$ as discussed in Section~\ref{sec:fluctuationsMagnetic} obeying equation~\eqref{eq:hxyMagnetic}.

\begin{figure}[]
\includegraphics[height=7cm]{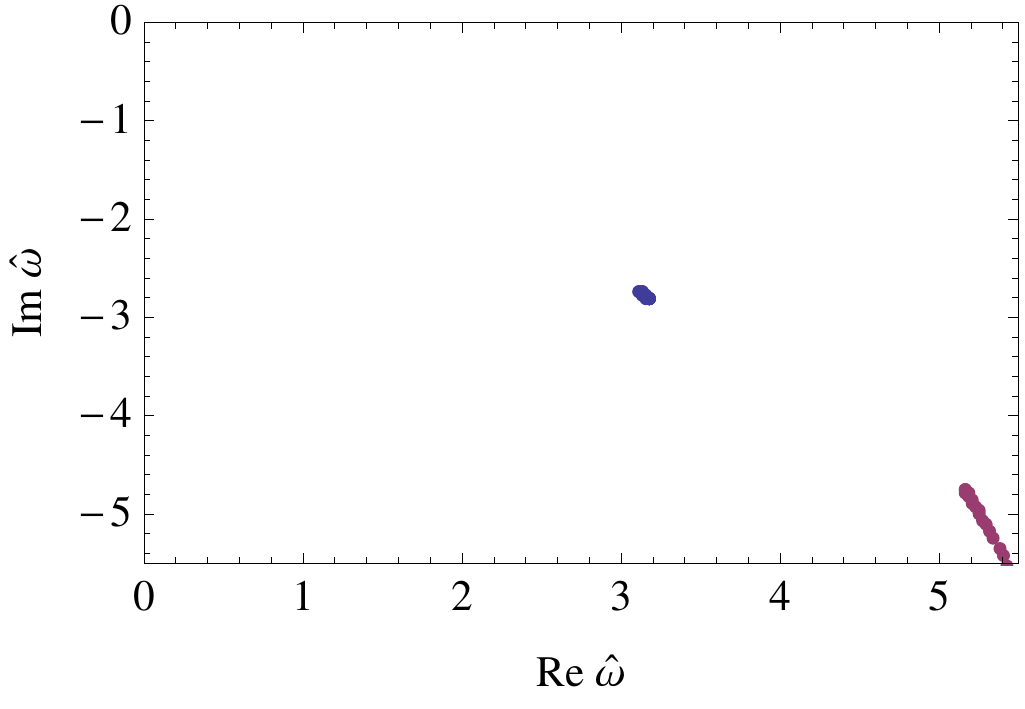}
\caption{\label{fig:scalarQNMsMagBrane}
{\it Trajectories of the two lowest QNMs of scalar ($h_{xx}-h_{zz})$ fluctuations around magnetic black branes:} The magnetic field is increased from $\mathcal{B}/T^2 = 0$ (at an energy density of $\epsilon/T^4 = 73.4$) to $15.5$ (at an energy density of $\epsilon/T^4 = 147.6$). These values are generated by fixing the near horizon parameter $u_1=2$, and choosing $b=0$ to $19/20$ in steps of $\Delta b=1/20$. Both, real and imaginary part increase when the magnetic field (and with it the energy density) is increased. The second QNM frequency moves a much greater distance in the complex frequency plane than the first.
}
\end{figure}

Let us begin with the scalar QNMs, see Fig.~\ref{fig:scalarQNMsMagBrane}. We provide a collection of sample values for the two lowest QNMs at three distinct points in $(T,\, \mathcal{B})$-parameter space, see table~\ref{tab:scalarQNMs_magBrane}.
\begin{table}[]
\begin{tabular}{|c|c|c|c|}
\hline
mode \#  & ${\mathcal{B}}/T^2=0$& ${\mathcal{B}}/T^2=12.953$  & ${\mathcal{B}}/T^2=30.161$  \\ \hline
1  & $\pm 3.1194506 -i\, 2.7466751$  & $\pm 3.4079677 - i\, 3.0777933$ & $\pm 3.8221728- i \, 3.5553514$   \\
2  & $\pm 5.1700747-i\, 4.7637826$  & $\pm 5.4444069-i\, 5.5805876$ & $\pm 5.6661971-i\, 6.6938458$   \\ \hline
\end{tabular}
\caption{ \label{tab:scalarQNMs_magBrane}
{\it Magnetic brane scalar QNMs.} The values of the two lowest lying quasinormal mode frequencies $\hat \omega$ for the scalar fluctuations, e.g. $h_{xx}-h_{zz}$, are shown in units of mass density. The columns show the quasinormal mode frequencies for three distinct values of $\mathcal{B}/T^2$. In the $\mathcal{B}\to 0$-limit, these two modes connect smoothly to the two lowest modes of set 1 of the RN black brane at vanishing charge density $\tilde q=0$, i.e. to the ones found earlier by Starinets~\cite{Starinets:2002br}.}
\end{table}
The lowest of the scalar QNMs in Tab.~\ref{tab:scalarQNMs_magBrane} is in reasonable agreement with the QNM values extracted from the late-time behavior in~\cite{Fuini:2015hba}, which are compared to our results in Tab.~\ref{tab:scalarQNMs_magBrane_compare}. The first column shows seven distinct values of $\mathcal{B}/T^2$; we confirmed that the (dimensionless) value of the energy density over $T^4$ agrees at these points within a few percent deviation between the non-equilibrium setup~\cite{Fuini:2015hba} and our magnetic brane thermodynamics from Sec.~\ref{sec:magBranes}. Agreement between the thermodynamic values for the two setups is best at small values of $\mathcal{B}/T^2$ (about 0.2 percent) at point 2, then deviations rise for bigger values, and asume their maximal value (about 2.3 percent) at point 7, i.e. for relatively large ${\mathcal{B}}/T^2=30.161$.
\begin{table}[]
\begin{tabular}{|l|c|c|c|}
\hline
parameter point & QNM & late-time~\cite{Fuini:2015hba} & relative deviation [in \%] \\
    &$\hat\omega
    $ & $\hat\omega
    $ & $\Delta \text{Re}\hat\omega  ,\, \Delta \text{Im}\hat\omega$ \\ \hline
1  $({\mathcal{B}}/T^2=0.000)$ & $\pm 3.119450641-i\,2.746675059$  & $\pm 3.1195 -i \, 2.7466$ &$\pm 0.00002 ,\, -0.00003$ \\
2  $({\mathcal{B}}/T^2=0.990)$  & $\pm 3.122662648-i\, 2.750477496$  & $\pm 3.124 - i\, 2.73 $ &  $\pm 0.04283, \, - 0.74451$  \\
3  $({\mathcal{B}}/T^2=5.344)$  & $\pm 3.197822654-i\, 2.838566372$  & $\pm 3.217 - i\, 2.79 $ &  $\pm 0.59970, \, - 1.71095$  \\
4  $({\mathcal{B}}/T^2=12.953)$  & $\pm 3.407967707-i\, 3.077793253$  & $\pm 3.480 - i \, 2.96 $ &  $\pm 2.11364 , \, - 3.8272$  \\
5  $({\mathcal{B}}/T^2=17.821)$  & $\pm 3.538016365- i \,3.224040043$  & $\pm3.634 - i\, 3.09 $ &  $\pm 2.71292 , \, - 4.15752$  \\
6  $({\mathcal{B}}/T^2=22.836)$  & $\pm 3.661199784-i\,3.364602583$  & $\pm 3.780 - i\,3.21  $ &  $\pm  3.24484 , \, - 4.59497$  \\
7  $({\mathcal{B}}/T^2=30.161)$  & $\pm 3.822172759 - i\, 3.555351392$  & $\pm 3.98 - i 3.38$ &  $\pm 4.12925, -4.93204$  \\ \hline
\end{tabular}
\caption{ \label{tab:scalarQNMs_magBrane_compare}
{\it Magnetic brane scalar QNMs versus far-from equilibrium late-time oscillations.} The value of the lowest lying quasinormal mode frequency $\hat \omega$ for the scalar fluctuations, e.g. $h_{xx}-h_{zz}$, are compared to the oscillation frequency extracted from the late-time behavior of the full time evolution in~\cite{Fuini:2015hba}. 
}
\end{table}

\begin{figure}[]
\includegraphics[height=7cm]{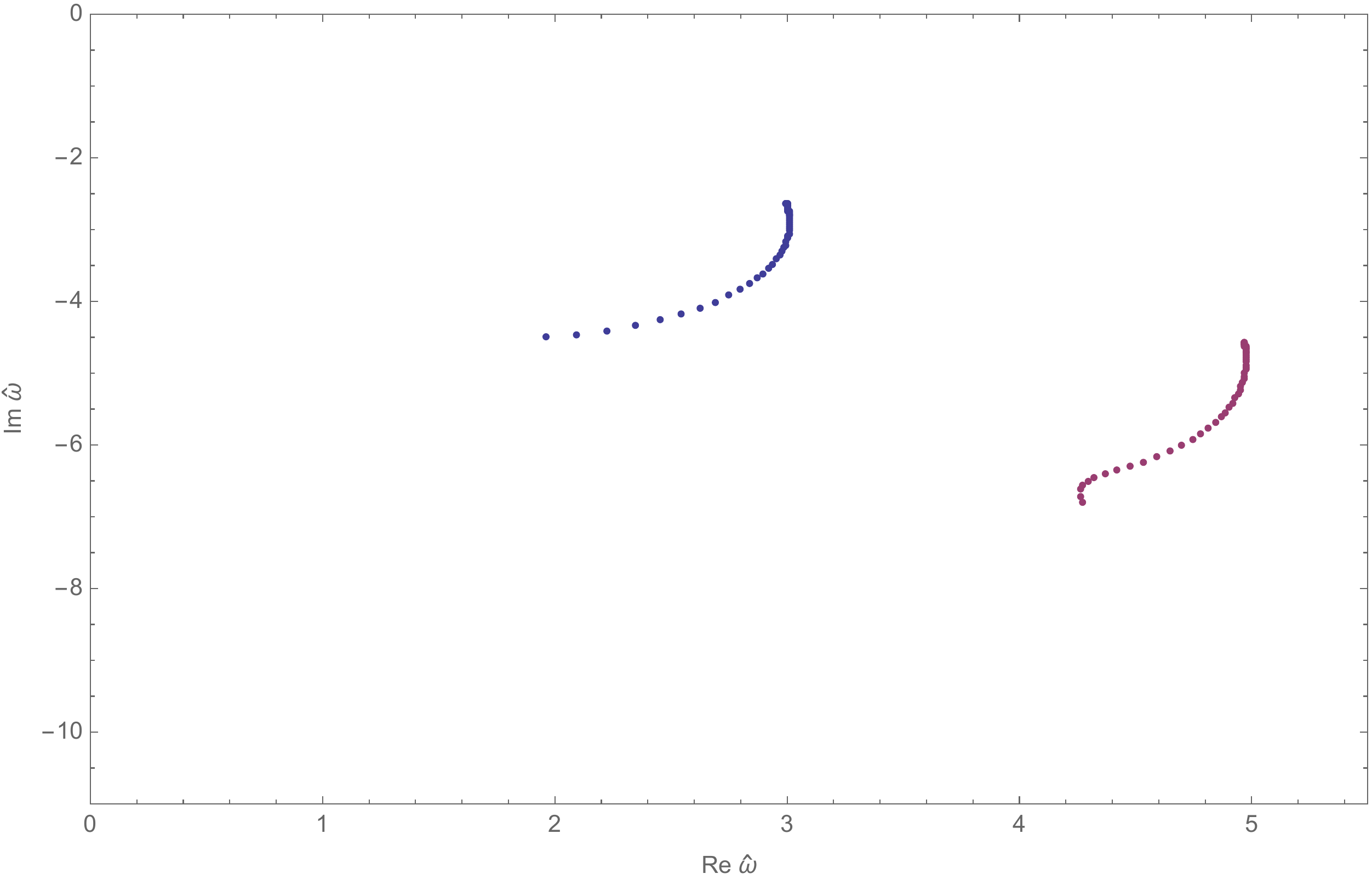}
\caption{\label{fig:QNMsMagBrane}
{\it QNMs of $h_{xy}$ fluctuations around magnetic black branes.} The two lowest QNMs (both members of set 1 in the limit of vanishing magnetic field, see Fig.~\ref{fig:QNMSets}) at vanishing momentum are shown as the magnetic field parameter $b$ is changed from $b=0$ to $b=50/40$ in steps of $\Delta b=1/40$. These modes both move down and to the left with increasing $b$. This covers a magnetic field over temperature range of $\mathcal{B}/T^2 = 0.0001,\, ..., 34.4555$, which corresponds to energy densities of $\epsilon/T^4= 73.4,\, ..., 397.4$. We have fixed the near horizon temperature parameter to $u_{1} = 2$. 
}
\end{figure}

Now we turn to the tensor QNMs at vanishing momentum, two of which are shown in Fig.~\ref{fig:QNMsMagBrane}. Note that the lowest tensor QNM approaches the imaginary frequency axis at large magnetic fields $\mathcal{B}\gg10$. For the second QNM, however, our numerical results are inconclusive at those large magnetic field values. We provide a collection of sample values for the two lowest QNMs (from set 1) at three distinct points in $(T,\, \mathcal{B})$-parameter space in Tab.~\ref{tab:tensorQNMs_magBrane}.
\begin{table}[]
\begin{tabular}{|c|c|c|c|}
\hline
mode \#  & ${\mathcal{B}}/T^2=5.342$& ${\mathcal{B}}/T^2=22.80$  & ${\mathcal{B}}/T^2=30.00$  \\ \hline
1  & $\pm 3.0104279 - i 2.8338442$  & $\pm 2.1829303 -i 3.4535515$ & $\pm 1.8555798, -i3.7287196$   \\
2  & $\pm 4.9749665- i 4.8923780$   &$\pm 3.6619677- i 5.3641251$ &  $\pm 3.6224530- i 5.5936925$ \\ \hline
\end{tabular}
\caption{ \label{tab:tensorQNMs_magBrane}
{\it Magnetic brane tensor QNMs.} The values of the two lowest lying quasinormal mode frequencies $\hat \omega$ for the tensor fluctuations, e.g. $h_{xy}$, are shown in units of mass density. The columns show the quasinormal mode frequencies for three distinct points in $(T, \, \mathcal{B})$ parameter space. In the $\mathcal{B}\to 0$-limit, these two modes connect smoothly to the two lowest modes of set 1 of the RN black brane at vanishing charge density $\tilde q=0$, i.e. to the ones found earlier by Starinets~\cite{Starinets:2002br}.}
\end{table}
%

%%%%%%%%%%%%%%%%%%%%%%%% C O N C L U S I O N S  
\section{Conclusion}\label{sec:conclusions}
Strongly coupled plasmas can be studied as they thermalize, because we have various holographic methods at our disposal. What is the correct picture for such an evolution from a ferocious state of matter to a tame thermal equilibrium? At late times in the thermalization process, we at least have a fairly clear picture on the gravity side: just like the overtones of a bell that was gently struck and then left to ring down freely, various quasinormal modes (QNMs) of the black brane are excited. All but the lowest lying QNMs (with the smallest imaginary part) have died down at late times because all but the lowest ones are strongly damped. This picture is valid for small deviations from equilibrium and at late times.
However, one may be inclined to think of intermediate states, and possibly even the initial state of the gravitational setup, as a superposition of (not necessarily quasinormal) modes.\footnote{A detailed analysis~\cite{Heller:2013oxa} found that a quasinormal mode decomposition approximates the full nonlinear time evolution surprisingly well for all tested initial conditions. In order to support the idea of formulating time-dependent systems in terms of a well-defined set of modes, we refer to the analysis of $AdS_5$ black hole QNMs performed in global $AdS$, and then reinterpreted in the Poincare patch~\cite{Friess:2006kw}. There, the authors relate that QNM spectrum to a time-dependent field theory stressing similarities to heavy ion collisions.} \footnote{Note also the perturbative weak field calculation of a gravitational collapse geometry in presence of charge~\cite{Caceres:2014pda}.}
Then, it would be interesting to know which of the modes are excited at what time during the evolution, what energy they carry, and how quickly they dissipate. Is all the energy first transferred from the low frequency modes into the high frequency modes, or vice versa? Our analysis in comparison with the far from equilibrium data from~\cite{Fuini:2015hba} allows a few speculations on which we comment below.

This present paper demonstrates the computation of tensor fluctuation ($h_{xy}$) quasinormal modes within two distinct black brane systems asymptoting to $AdS_5$: First, Reissner-Nordstr\"om (charged) black branes. Second, magnetic black branes (uncharged). In the latter, we also compute the QNMs of scalar fluctuations $h_{tt},\, h_{xx},\, h_{yy}, \, h_{zz}$, 
while in the Reissner-Nordstr\"om case we re-compute the vector QNMs originally found in~\cite{Maeda:2006by}.
We provide sample values in tables \ref{tab:QNMsRN_consistency}, \ref{tab:QNMsRN_k0}, \ref{tab:QNMsRN_khalf}, \ref{tab:QNMsRN_kone}, \ref{tab:vectorQNMs}, \ref{tab:scalarQNMs_magBrane}, \ref{tab:scalarQNMs_magBrane_compare},\ref{tab:tensorQNMs_magBrane}. Two notebooks are provided online~\cite{notebookLink} for computation of QNMs at user-selected charge density $\tilde q$ and momentum $k_z$, or for reference QNM frequency values at particular magnetic fields $\mathcal{B}$, and temperatures $T$. All these quasinormal mode frequencies should be used in order to compare to holographic calculations of anisotropic fluids far from equilibrium. The late-time behavior of the latter should be governed by oscillations and decay that is well approximated by particular QNMs. 

As one example, we have compared our QNMs to a system~\cite{Fuini:2015hba} which starts with a metric that is anisotropic between the $xy$-plane, and the $z$-direction, and which is then observed as it freely evolves either in the presence of (i) charges or (ii) a magnetic field on the black brane.
We find good agreement. In the case of a charged plasma, dual to the Reissner-Nordstr\"om $AdS_5$ black brane (case (i)), our lowest tensor mode QNM (associated with fluctuations $h_{xy}$ or equivalently $h_{xx}-h_{zz}$) agree with the values provided in~\cite{Fuini:2015hba} up to 0.2\% deviation at large charge densities.\footnote{This may have to do with the lowest mode of set 2, beginning to play a significant role at these larger densities; but not yet dominating the late-time behavior.} In the case of a magnetic plasma, dual to the magnetic $AdS_5$ black brane case (case (ii)), our lowest scalar QNM (associated with fluctuations $h_{xx} - h_{yy}$) is in good (at worst 5\% deviation) agreement with~\cite{Fuini:2015hba}, see Tab.~\ref{tab:scalarQNMs_magBrane}.
These findings confirm our reasoning from Section~\ref{sec:FFEnQNM}, that the late-time behavior of a relaxing system is determined by the symmetries of the background it relaxes towards, and by the symmetries of the initial conditions. In the Reissner-Nordstr\"om case, the initial shear in the setup~\cite{Fuini:2015hba} (far from equilibrium) is a tensor under $SO(2)$ rotations in the $xy$-plane. At late times, only the corresponding tensor QNM matters. In the magnetic black brane case, on the other hand, the same initial shear is a scalar. At late times, only the corresponding scalar QNMs matter.
 
Furthermore, we provide QNM values for Reissner-Nordstr\"om black branes at nonzero momentum. We may think of these QNMs as being useful for comparison to far from equilibrium systems which started out with an initial condition depending on the $z$-direction. Colliding shock waves with a $z$-dependent profile may serve as an example for this kind of system. 
 
At sizable charge densities, a novel set of QNMs dominates the late-time behavior of the charged black brane system, and hence also the late-time behavior of the dual charged plasma. In addition to the well known set of QNMs (set 1) of the Reissner-Nordstr\"om black brane, we also found a set of purely imaginary QNMs (set 2) as schematically shown in Fig.~\ref{fig:QNMSets}. 
These modes enter the low energy spectrum at and above a particular charge density $\tilde q\ge 0.655$, corresponding to a sizable temperature $T \le T_{crossing} \approx 0.228$, with the lowest member of that set becoming the new lowest lying QNM of the sytem. As the sytem is pushed toward extremality, based on our numerics, we conjecture that these modes of set 2 are going to be pushed towards each other and form a branch cut. An analogous behavior is seen in the $AdS_4$ case~\cite{Edalati:2010hk,Edalati:2010pn}. 
 
In our second system, the magnetic black branes, we also find tensor QNMs at vanishing momentum. We provide sample values in table~\ref{tab:tensorQNMs_magBrane}, and provide a notebook with samples of QNMs at various temperatures, and magnetic fields.

It is interesting to note that all of the QNMs of magnetic branes move deeper into the complex frequency plane as the magnetic field is increased. This should be interpreted as a faster dissipation of each mode after it has been excited. It is then tempting to argue for a faster equilibration of the non-equilibrium system if all the QNMs decay faster at larger magnetic fields. However, this does not seem to be the case~\cite{Fuini:2015hba}. Instead, (charge or) magnetic field individually seem to have a rather negligible effect on the equilibration time~\cite{Fuini:2015hba}. This may be interpreted in (at least) two ways: as evidence for the fact that 
\begin{enumerate}
\item[(i)] the standard QNMs (which we computed here) are not a good measure for equilibration of this system while it is far from equilibrium, and one needs to find other ways of characterizing this process, or 
\item[(ii)] only the lowest of the QNMs plays a significant role in the equilibration process. This is a possibility, because the lowest magnetic brane QNM remains virtually static, while the higher magnetic brane QNMs change significantly with the magnetic field, as illustrated in Fig.~\ref{fig:scalarQNMsMagBrane}.
\end{enumerate}

We point out a fundamentally different behavior for a subset of the electric black brane QNMs. As noted in the previous paragraph for magnetic brane QNMs, most of the electric brane QNMs move deeper into the complex plane at increasing charge density. The noteworthy exception are the QNMs of set 2, see Fig.~\ref{fig:QNMSets}. Those QNMs move closer to the origin of the complex frequency plane while the charge is increased (and simultaneously temperature approaches zero).~\footnote{Numerically, we can only verify this up to a certain lowest charge value, at which these imaginary QNMs are simply too deep in the complex frequency for our methods to find them at the current accuracy. However, all available results indicate that these imaginary QNMs move closer to $\omega=0$ with increasing charge, at any given charge value.} This behavior indicates a decreased dissipation of these modes, i.e. these modes become more and more long-lived. It is tempting to ask if these modes become part of an effective field theory description near zero temperature. Such a description could contain these long-lived modes whose frequency is much smaller than the charge density, i.e. $\omega\ll \tilde q$, near zero temperature. This would be in anology to the standard hydrodynamic description of long-lived modes which satisfy the relation $\omega\ll T$. Such an effective field theory description may be useful for the study of black holes near extremality. It should be stressed that these modes are not propagating.
 
To provide a brief outlook, it would be interesting to consider the other fluctuations of our two systems. All fluctuations could be studied at nonzero momentum in the $z$-direction. The same can be done for a nonzero momentum in the $y$-direction. In that case, one would then compute the QNMs of metric fluctuations which are vectors or scalars under the remaining rotation group. Among those QNMs one expects diffusion and sound modes, i.e. modes with dispersion relation $\omega \propto k^2$ and $\omega\propto k$, respectively. This would then allow one to compare to the late-time behavior of more complicated far from equilibrium systems. 

Finally, we note that the shooting and continued fraction methods take quite a long time to find the present results. More realistic, and thereby more complicated systems, do encourage the development of improved methods for the computation of QNMs.

%%%%%%%%%%%%%%%%%%%%%%%% A C K N O W L E D G M E N T S
\section*{Acknowledgements}
We thank J.~Fuini, L.~Yaffe for many helpful and inspiring discussions. We are grateful to M.~Heller and J.~Sonner for discussions and for sharing some of their code. We thank A.~G.~Green, A.~Karch, W.~van~der~Schee, and J.~Zaanen for discussions. This work was supported in part by the US Department of Energy under contract number DE-SC0011637, and in part by NSERC of Canada.

%%%%%%%%%%%%%%%%%%%%%%%% A P P E N D I X
\begin{appendix}
%%%%%%%%%%%%%%%%%%%%%%%%%%%%%%%%%%%%%%%
\section{Generalization of Pincherle's theorem and continued fraction method}\label{app:Pincherle}

In its generalization, Pincherle's theorem relates $m+2$ term recursion relations to $m$-dimensional continued fractions. Such vectorial continued fractions require use of a projective space, all needed notions will be defined as they arise. The first useful result is Parusnikov's Theorem 2 which provides the correspondence
\newpage
\begin{center}
\textbf{Theorem 2:}\\
i) interuption-free $m$-dimensional continued fraction:\\
$\vec{f}_0=\vec{b}_0+\frac{a_0}{\vec{b}_1+\frac{a_1}{\vec{b}_2+\cdots}}$,\\
$\iff$\\
iii) $m+2$ term recursion relation: $g_n=\sum\limits^{m+1}_{j=1}p_{m+2-j,n}g_{n-j}$,
\end{center}
where: $\vec{p}_n\equiv(1/a_n,b_{1,n}/a_n,\cdots,b_{m,n}/a_n)$; generically $v_{i,j}$ denotes the $i$-th component of vector $\vec{v}_j$;  and ``interuption-free'' means that $a_n$ is non-zero.

The construction of the $m$-dimensional continued fraction is at first opaque, as it requires ``dividing by a vector,'' but it can be simply accomplished by the following procedure. First, given the vector $\vec{p}_n$, define
\begin{align}
\textbf{P}_n\equiv\begin{pmatrix}
0	& \cdots & 0 & p_{1,n} \\
1	& \cdots & 0 & p_{2,n} \\
\vdots & \ddots & \vdots & \vdots \\
0	& \cdots & 1 & p_{m+1,n}
\end{pmatrix},\quad \textbf{Q}_n=\textbf{P}_0\cdots \textbf{P}_n,\quad\textbf{J}^{-1}\equiv \begin{pmatrix}
	0&1&0&\cdots&0\\
	0&0&1&\cdots&0\\
	\vdots&\vdots&\vdots&\ddots&\vdots\\
	0&0&0&\cdots&1\\
	1&0&0&\cdots&0\\
\end{pmatrix}.
\end{align}

These matrices act on $m+1$-dimensional projective vectors, denoted $(f_1:\cdots:f_{m+1})$. For the map $\sigma^{-1}\left[(f_1:\cdots:f_{m+1})\right]\equiv(f_1/f_{m+1},\cdots,f_m/f_{m+1})$, the $m$-dimensional continued fraction can be computed
\begin{equation}
\vec{f}_0\equiv\lim_{i\to\infty}\vec{f}_0^{\ (i)}\equiv\lim_{i\to\infty}\sigma^{-1}\textbf{J}^{-1}\textbf{Q}_i\vec{e}_{m+1},
\label{eq:f0}
\end{equation}
where $\vec{e}_{m+1}\equiv(0:\cdots:0:1)$. Theorem 2 therefore provides a way to construct an $m$-dimensional continued fraction given an $m+2$ term recursion relation.

The next theorem provides a powerful criterion for when the continued fraction $\vec{f}_0$ converges.
\begin{center}
\textbf{Theorem 6:}\\
An $m$-dimensional continued fraction $\vec{f}_0$ converges.\\
$\iff$\\
There exists a minimal sequence $h_n$ that satisfies the recursion relation given in Theorem 2.
\end{center}

Parusnikov's Theorems 7 and 8 now give a recursion relation that allows an explicit determination of $h_n$.
\begin{center}
\textbf{Theorem 7/8:}\\
The minimal basis $h_n$ obeys:\\
\begin{equation}
h_n=\left(\frac{-1}{f_{m,n+1}}\right)h_{n-m}+\sum\limits^{m-1}_{j=1}\left(\frac{-f_{j,n+1}}{f_{m,n+1}}\right)h_{n-m+j}\, ,
\end{equation}
\end{center}
where $\vec{f}_n$ is the continued fraction $\vec{f}_n\equiv \vec{b}_n+a_n/(\vec{b}_{n+1}+\cdots)$. This formula allows the calculation of the quasinormal frequencies, as explicitly demonstrated in section \ref{sec:ex}. By setting $n=m$, the $h_i$ become the first $m+1$ coefficients of the recursion relation generated by the equation of motion. One then uses \eqref{eq:f0} to plug in the components of $\vec{f}_{m+1}$ as a function of $\tilde\omega$. Demanding the veracity of the formula in Theorem 7/8 determines the frequencies that give the minimal solution, and hence are the quasinormal frequencies, as discussed above in section \ref{sec:pin}.

%%%%%%%%%%%%%%%%%%%%%%%%%%%%%%%%%%%%%%%%%%%%%%%%%%%%%%%%%%%

\end{appendix}

%%%%%%%%%%%%%%%%%%%%%%%% R E F E R E N C E S
%\bibliographystyle{plain}     % bibtex style file 
%\bibliographystyle{revtex4-1}
%\bibliographystyle{h-physrev}

\end{document}